%% file: security_arxiv.tex
\documentclass[onecolumn]{IEEEtran}
\input{preamble}

\usepackage[ colorlinks = true,
             linkcolor = blue,
             urlcolor  = blue,
             citecolor = red,
             anchorcolor = green,]{hyperref}
\usepackage{cite}
\usepackage{extarrows}
\usepackage{ulem}
\begin{document}

\title{Multiple Private Key Generation for Continuous Memoryless Sources with A Helper}
\author{\IEEEauthorblockN{Lin Zhou}
\thanks{Lin Zhou was with the Department of Electrical Engineering and Computer Science at the University of Michigan, Ann Arbor. He is now with School of Cyber Science and Technology, Beihang University, Beijing, China (Email: lzhou@buaa.edu.cn).}
}
\maketitle

\begin{abstract}

We propose a method to study the secrecy constraints in key generation problems where side information might be present at untrusted users. Our method is inspired by a recent work of Hayashi and Tan who used the R\'enyi divergence as the secrecy measure to study the output statistics of applying hash functions to a random sequence. By generalizing the achievability result of Hayashi and Tan to the multi-terminal case, we obtain the output statistics of applying hash functions to multiple random sequences, which turn out to be an important tool in the achievability proof of strong secrecy capacity regions of key generation problems with side information at untrusted users. To illustrate the power of our method, we derive the capacity region of the multiple private key generation problem with an untrusted helper for continuous memoryless sources under Markov conditions. The converse proof of our result follows by generalizing a result of Nitinawarat and Narayan to the case with side information at untrusted users.
\end{abstract}

\section{Introduction}
The problem of generating a secret key for two parties observing correlated random variables was first considered by Maurer~\cite{maurer1993} and by Ahlswede and Csisz\'ar~\cite{ahlswede1993}. In \cite{maurer1993,ahlswede1993}, there are two legitimate users Alice and Bob as well as an eavesdropper Eve. Alice observes a source sequence $A_1^n$, Bob observes $A_2^n$ and Eve observes $A_3^n$. It is assumed that there exists noiseless public channel over which Alice and Bob can talk interactively in $r$ rounds. The eavesdropper, although not allowed to talk, can overhear the messages transmitted over the public channel. Under the condition that $A_1-A_3-A_2$ forms a Markov chain, it is shown that the secret key capacity (maximal rate of the secret key) is $I(A_1;A_2|A_3)$ for discrete memoryless sources (DMS). 

Subsequently in \cite{csiszar2000}, Csisz\'ar and Narayan extended the model in \cite{maurer1993,ahlswede1993} by adding a third party called a helper which assists the legitimate users to generate a secret key. Furthermore, the authors in \cite{csiszar2004secrecy} generalized the result in \cite{csiszar2000} to a setting with at least four terminals. It is assumed in~\cite{csiszar2004secrecy} that there exist an eavesdropper Eve and other $T\geq 3$ terminals denoted by $\calT$. For each $t\in\calT$, terminal $t$ observes a source sequence $A_t^n$, which is correlated with all other source sequences $\{A_i^n\}_{i\in\calT:i\neq t}$. The eavesdropper observes a correlated source sequence $E^n$. Let $\calS$ and $\calW$ denote two disjoint group of users, i.e., $\calS\bigcap\calW=\emptyset$. All users in $\calS$ aim to generate a common key $K$ with the help of all other users in $\calT$. Interactive communication with unlimited rate is assumed and the overall communication over the public channel is denoted as $\bF$. The authors in \cite{csiszar2004secrecy} considered three problems depending the security constraint on the key. If the key is only concealed from the public messages $\bF$, the problem is a secret key generation problem. If the key is concealed from both the public messages $\bF$ and the source sequences observed by users $\calW$, then the problem is a private key generation problem. If the key is concealed from both the public messages $\bF$ and the source sequence observed by the eavesdropper, the problem is considered as a wiretap key generation problem. Using results in the distributed source coding~\cite[Theorem 3.1.14]{csiszar2011information}, Csisz\'ar and Narayan~\cite{csiszar2004secrecy} characterizes the exact capacity for the secret key and private key generation problems as well as bounds on the wiretap key capacity. Furthermore, the authors proved an upper bound on the secret key capacity and conjectured the upper bound is tight in general. The conjecture was solved partially by Ye and Reznik~\cite{ye2007} and proved to be true by Chan and Zheng~\cite{chan2010}. Other works on the secret key generation include \cite{tyagi2015,hayashi2016ssecond,narayan2008,hayashi2011,chou2015,khisti2011,bloch2011physical,chou2014separation}.

The problem of generating multiple keys was initialized by Ye and Narayan in \cite{ye2005isit} where they considered the generation of a private key and a secret key with three terminals. The authors proved an outer bound on the capacity region which was later shown to be tight by Zhang \emph{et al.}~\cite{zhang2014}. Furthermore, in \cite{zhang2017}, the authors considered generating two keys in a cellular model and derived the capacity region for four cases depending on the security constraints on the keys. Other works on the multiple key generation problem include \cite{tu2017mkey,ye2012,xu2016}. In terms of key generation problems for correlated Gaussian memoryless sources (GMS), Nitinawarat and Narayan~\cite{narayan2012} derived the capacity for secret key generation with multi-terminals and thus extended \cite[Theorem 2]{csiszar2004secrecy} to GMS. Watanabe and Oohama derived the capacity for secret key generation for GMS and vector GMS under rate-limited public communication in \cite{watanabe2010secret} and \cite{watanabe2011secret} respectively. Other works on secret key generation for GMS include \cite{ye2006ISIT,khisti2016,khisti2012}.

In this paper, we are interested in the private key generation problem for correlated continuous memoryless sources (CMS) with a helper and unlimited public discussion. To the best of our knowledge, the private key generation problem for CMS remains unexplored. 

The main challenges of private key generation problems for CMS lie in the analysis of the secrecy constraints in the achievability part since we need to upper bound the term $I(K;A^n,\bF)$ where $K$ is the private key, $\bF$ is the public message and $A^n$ is an continuous i.i.d. sequence observed by some untrusted helpers. To bound $I(K;A^n,\bF)$, existing works, e.g., \cite{khisti2016,khisti2012,chou2014separation}, applied quantization to the continuous side information $A^n$ and relied heavily on the continuity of information quantities. The analyses are usually tedious.

In contrast, inspired by \cite{hayashi2011} and \cite{hayashi2017}, we analyze the secrecy part in the private key generation problem for CMS by studying the output statistics of hash functions (random binning) under the R\'enyi divergence measure and using the fact that the R\'enyi divergence is non-decreasing in the order~\cite{van2014renyi}. The great advantage of our proposed method is that it is a unified and neat method which holds for the case with either continuous, discrete or no side information at untrusted users. We believe that our result in Lemma \ref{propmeasure} can be used to significantly simplify the security analysis for secret key generation problems when the eavesdropper has access to continuous (e.g., Gaussian) side information which are correlated to the observations at legitimate users~(e.g., \cite{khisti2016,khisti2012,chou2014separation}). Furthermore, our proposed method can be used to derive bounds on the convergence speed of secrecy constraints beyond the fact that secrecy constraints vanish under certain rate constraints. See the remark after Theorem \ref{strongcapacity} for further discussion.

\subsection{Main Contributions}
\label{sec:maincontri}
Our main contributions are summarized as follows.

Firstly, we derive the output statistics of applying hash functions to multiple random sequences under the R\'enyi divergence measure in Lemma \ref{propmeasure}. Lemma \ref{propmeasure} is an extension of \cite[Theorem 1]{hayashi2011} to a multi-terminal case and a strict generalization of \cite[Theorem 1]{yassaee2014} where the output statistics of random binning under the total variational distance measure was derived. Furthermore, Lemma \ref{propmeasure} turns out to be an important tool in analyzing secrecy constraints in key generation problems, especially when the key needs to be protected from continuous observations correlated to observations at legitimate users.

Secondly, to illustrate the power of Lemma \ref{propmeasure}, we derive the capacity region for the multiple private key generation problem with a helper for CMS. To be specific, we revisit the model in~\cite{zhang2017} and derive the capacity region for CMS under a symmetric security requirement which did not appear in \cite{zhang2017}. The converse proof follows by judiciously adapting the techniques in \cite[Theorem 1]{narayan2012} to our setting. In the achievability proof, we use Lemma \ref{propmeasure} to analyze the secrecy constraints on generated keys which need to be protected from correlated continuous observations of illegitimate users. Furthermore, we use the quantization techniques in \cite{narayan2012}, the large deviations analysis for distributed source coding in \cite{gallager}, the Fourier Motzkin Elimination and the techniques to bound the difference between the differential entropy of CMS and the discrete entropy of the quantized random variables. We remark that the techniques used in our paper can also apply to strengthen all the four cases in \cite{zhang2017} with strong secrecy and for CMS. Furthermore, we also extend our result to a cellular model involving more than four terminals and derive inner and outer bounds for the capacity region.

\subsection{Organization of the Paper}
The rest of the paper is organized as follows. In Section \ref{sec:notation}, we set up the notation. In Section \ref{sec:oshf}, we formulate the problem of output statistics of hash functions and present our main result under the R\'enyi divergence measure in Lemma \ref{propmeasure}. Subsequently in Section \ref{sec:capacityresult}, invoking Lemma \ref{propmeasure}, we derive the capacity region for the multiple private key generation problem with a helper. Furthermore, we generalize our result to a cellular model and derive bounds on the capacity region. The proofs of the capacity region for the multiple private key generation with a helper are given Sections \ref{proofstrongc} and \ref{prooftight}. Finally, we conclude our paper and discuss future research directions in Section \ref{sec:conc}. For the smooth presentation of our main results, the proofs of all supporting lemmas are deferred to the appendices.

\subsection*{Notation}
\label{sec:notation}
Throughout the paper, random variables and their realizations are in capital (e.g.,\ $X$) and lower case (e.g.,\ $x$) respectively. All sets are denoted in calligraphic font (e.g.,\ $\mathcal{X}$). We use $\calX^{\mathrm{c}}$ to denote the complement of $\calX$ and use $U_{\calX}$ to denote the uniform distribution over $\calX$. Given any two integers $a,b$, we use $[a:b]$ to denote the set of all integers between $a$ and $b$ and we use $[a]$ to denote $[1:a]$ for any integer $a\geq 1$. Let $\calT:=\{1,\ldots,T\}$. Given a sequence of random variables $X_1,X_2,\ldots,X_T$ and any subset $\calS\subseteq\calT$, we use $X_{\calS}$ and $\{X_t\}_{t\in\calS}$ interchangeably. Furthermore, let $X^n:=(X_1,\ldots,X_n)$ be a random vector of length $n$.  For any $(a,b)\in[1:n]^2$, we use $X_a^b$ and $(X_a,\ldots,X_b)$ interchangeably. For information theoretical quantities, we follow~\cite{cover2012elements}.

\section{Output Statistics of Hash Functions}
\label{sec:oshf}
In this subsection, we consider hash functions and study its output statistics under the R\'enyi divergence measure. The result in this section (cf. Lemma \ref{propmeasure}) serves as an important tool in the subsequent analysis for key generation problems.

\subsection{Preliminary}
Before presenting the main result, we first introduce some definitions. Given two distributions $(P_{A_1},Q_{A_1})$ defined an alphabet $\calA_1$, the KL divergence is defines as 
\begin{align}
D(P_{A_1}\|Q_{A_1})
&:=\sum_{a_1\in\calA_1}P_{A_1}(a_1)\log\frac{P_{A_1}(a_1)}{Q_{A_1}(a_1)}.
\end{align}
Furthermore, given $s\in[-1,\infty)$, the R\'enyi divergence or order $1+s$ is defined as
\begin{align}
D_{1+s}(P_{A_1}\|Q_{A_1})&:=
\left\{
\begin{array}{ll}
\frac{1}{s}\log\sum_{a_1\in\calA_1}P_{A_1}^{1+s}(a_1)Q_{A_1}^{-s}(a_1)&s\neq 0,\\
D(P_{A_1}\|Q_{A_1})&s=0.
\end{array}
\right.\label{def:rydivergence}
\end{align}
It is well known that $D_{1+s}(P_{A_1}\|Q_{A_1})$ is non-decreasing in $s$ (cf.~\cite{van2014renyi}) and thus $D(P_{A_1}\|Q_{A_1})\leq D_{1+s}(P_{A_1}\|Q_{A_1})$ for all $s\geq 0$. 

Given a joint distribution $P_{A_1E}$ on the alphabet $\calA_1\times\calE$, the conditional entropy is defined as 
\begin{align}
H(A_1|E)
&:=-\sum_{e\in\calE}P_E(e)\sum_{a_1\in\calA_1}P_{A_1|E}(a_1|e)\log P_{A_1|E}(a_1|e).
\end{align}
Furthermore, given $s\in[-1,\infty)$, the conditional R\'enyi entropy of order $1+s$ is defined as
\begin{align}
H_{1+s}(A_1|E)
&:=
\left\{
\begin{array}{ll}
-\frac{1}{s}\log\sum_e P_E(e)\sum_{a_1} P_{A_1|E}^{1+s}(a_1|e)&s\neq 0,\\
H(A_1|E)&s=0,
\end{array}
\right.\label{def:condryentropy}
\end{align}
and the Gallager's conditional R\'enyi entropy of order $s$ is defined as
\begin{align}
H_{1+s}^{\uparrow}(A_1|E)&:=-\frac{1+s}{s}\log \sum_e \Big(\sum_{a_1} P_{A_1E}^{1+s}(a_1,e)\Big)^{\frac{1}{1+s}}\label{def:condentropyG}.
\end{align}
We remark that for continuous random variables, the summations in \eqref{def:rydivergence}, \eqref{def:condryentropy}, \eqref{def:condentropyG} should be replaced by integrals.

We then recall the formal definition of a hash function \cite[Eq. (1)]{tsurumaru2013dual} (see also~\cite{wegman1981new,carter1979universal}).

\begin{definition}
Given an arbitrary set $\calA$ and the set $\calM:=\{1,\ldots,M\}$, a random hash function $f_X$ is a stochastic mapping from $\calA$ to $\calM$, where $X$ denotes the random variable describing the stochastic behavior of the hash function. Given any $\varepsilon\in\bbR_+$, an ensemble of random hash functions $f_X$ is called an $\varepsilon$-almost universal$_2$ hash function if it satisfies that for any distinct $(a_1,a_2)\in\calA^2$, we have 
\begin{align}
\Pr\{f_X(a_1)=f_X(a_2)\}\leq \frac{\varepsilon}{M}.
\end{align}
When $\varepsilon=1$, we say that the ensemble of functions is a universal$_2$ hash function.
\end{definition}
We remark that random binning in source coding problems~(e.g.,~\cite[Chapter 15.4.1]{cover2012elements}) is a universal$_2$ hash function. 

\subsection{Output Statistics}
In this subsection, we study the output statistic of applying hash functions to multiple random sequences under the R\'enyi divergence measure. For simplicity, we use $\calT$ to denote the set $\{1,\ldots,T\}$. Consider a sequence of random variables $(A_{\calT},E)=(A_1,\ldots,A_T,E)$ with a joint distribution $P_{A_\calT E}$ defined on an alphabet $\prod_{t\in\calT}\calA_t\times\calE$ where all $t\in\calT$, the alphabet $\calA_t$ is finite. Let $(A_{\calT}^n,E^n)$ be an i.i.d sequence generated according to the distribution $P_{A_\calT E}$. 

For each $t\in\calT$, let $f_{X_t}^{(n)}$ be an $\varepsilon$-almost universal$_2$ hash function mapping from $\calA_t^n$ to $\calM_t:=\{1,\ldots,N_t\}$ where $X_t$ describes the stochastic behavior of the hash function. Furthermore, the rate of the hash function $f_{X_t}$ is defined as $R_t:=\frac{1}{n}\log N_t$. We are interested in the output statistics of applying hash functions to the random sequences $A_\calT^n$, i.e., $\{f_{X_t}^{(n)}(A_t^n)\}_{t\in\calT}$.

For ease of notation, let $M_t:=f_{X_t}^{(n)}(A_t^n)$ for each $t\in\calT$. Furthermore, for each $t\in\calT$, let $U_{\calM_t}$ denote the uniform distribution over $\calM_t$ and let $P_{M_t}$ denotes the induced output distribution by $P_{A_t}^n$ and the hash function $f_{X_t}$, i.e., for all $m_t\in\calM_t$,
\begin{align}
P_{M_t}(m_t)=\sum_{a_t^n\in\calA_t^n}P_{A_t}^n(a_t^n)1\{f_{X_t}(a_t^n)=m_t\}.
\end{align}
To quantify the output statistics of the hash functions, it is common to use the KL divergence $D(P_{M_{\calT}E^n}\|\prod_{t\in\calT}U_{\calM_t}\times P_E^n)$ as a measure where
\begin{align}
D(P_{M_{\calT}E^n}\|\prod_{t\in\calT}U_{\calM_t}\times P_E^n)
&=D(P_{M_{\calT}}\|\prod_{t\in\calT}U_{\calM_t})+I(M_{\calT};E^n)\\
&=\sum_{t\in\calT:t\geq 2} I(M_t;M_{[1:t-1]})+\sum_{t\in\calT}D(P_{M_t}\|U_{\calM_t})+I(M_{\calT};E^n)\label{secrecym1}.
\end{align}
Note that if $D(P_{M_{\calT}E^n}\| \prod_{t\in\calT}U_{\calM_t}\times P_E^n)<\delta$ for some $\delta>0$, then we have the following results
\begin{enumerate}
\item $\sum_{t\in\calT:t\geq 2}^T I(M_t;M_{[1:t-1]})$ is small, indicating that the output of hash functions $M_{t_1}$ and $M_{t_2}^n$ are almost independent for all distinct pairs $(t_1,t_2)\in\calT^2$; 
\item $\sum_{t\in\calT} D(P_{M_t}\|U_{\calM_t})$ is small, indicating that the output of each hash function $M_t$ is almost uniform over $\calM_t$ for all $t\in\calM_t$; 
\item $I(M_{\calT};E^n)$ is small, indicating that the collection of outputs of hash functions $M_{\calT}=(M_1,\ldots,M_T)$ is almost independent of the side information $E^n$. 
\end{enumerate}

In this subsection, instead of using \eqref{secrecym1}, we make use of the R\'enyi divergence of order $1+s$ (cf. \eqref{def:rydivergence}) as the measure of output statistics of hash functions, i.e.,
\begin{align}
C_{1+s}(M_{\calT}|E^n)
&:=D_{1+s}(P_{M_{\calT}E^n}\|\prod_{t\in\calT}U_{\calM_t}\times P_E^n)\\
&=\sum_{t\in\calT}\log M_t-H_{1+s}(M_{\calT}|E^n)\label{secrecym2},
\end{align}
where $s\in(0,1]$ \eqref{secrecym2} follows from \eqref{def:condryentropy}. Note that the measure in \eqref{secrecym2} is a strict generalization of that in \eqref{secrecym1}.

Our results in the following lemma concern the output statistics of $\varepsilon$-almost universal$_2$ hash functions for any $\varepsilon\in\bbR_+$ unless otherwise stated.
\begin{lemma}
\label{propmeasure}
The following claims hold.
\begin{enumerate}
\item For any $s\in[0,1]$
\begin{align}
\mathbb{E}_{X_{\calT}}\Big[\exp(sC_{1+s}(M_{\calT}|E^n))\Big]\leq \varepsilon^{sT}+\sum_{\emptyset\neq\calS\subseteq\calT} \varepsilon^{s(T-|\calS|)}\Big(\prod_{i\in\calS} N_t^s\Big)\exp(-sn H_{1+s}(A_{\calS}|E))\label{oneshot}.
\end{align}
\item For any $s\in[0,1]$, if for all non-empty subset $\calS$ of $\calT$,
\begin{align}
\sum_{t\in\calS}R_t<H_{1+s}(A_{\calS}|E)\label{secrecycondition}
\end{align}
then
\begin{align}
\lim_{n\to\infty}\frac{1}{n}\mathbb{E}_{X_{\calT}}\Big[C_{1+s}(M_{\calT}|E^n)\Big]
&=0\label{weaksecrecy};
\end{align}
\item When $\varepsilon=1$, for any $s\in[0,1]$
\begin{align}
\liminf_{n\to\infty}-\frac{1}{n}\log \mathbb{E}_{X_{\calT}}[C_{1+s}(M_{\calT}|E^n)]&\geq \max_{\theta\in[s,1]}\min_{\emptyset\neq\calS\subseteq\calT} \theta(H_{1+\theta}(A_{\calS}|E)-\sum_{t\in\calS}R_t)\label{strongsecrecy}.
\end{align}
\end{enumerate}
\end{lemma}
Note that the asymptotic performance in Claim (iii) is achieved only by ($1$-almost) universal$_2$ hash functions since we put the additional constraint of $\varepsilon=1$. This condition could potentially be relaxed with techniques in~\cite{hayashi2016sa}.

The proof of Lemma \ref{propmeasure} is inspired by \cite[Theorem 1]{hayashi2011}, \cite[Lemma 1 and Theorem 2]{hayashi2017} and provided in Appendix \ref{proofpropmeasure}. A few remarks are in order. 

Firstly, Lemma \ref{propmeasure} is a generalization of \cite[Theorem 1]{hayashi2011} to multi-terminal. In the proof of Lemma \ref{propmeasure}, we also borrow an idea from \cite[Theorem 1]{yassaee2014} which studied the output statistics of universal$_2$ hash functions under the total variation distance measure instead of the R\'enyi divergence considered here. Invoking \eqref{basis4proof} and Pinsker's inequality, it is easy to see that that \cite[Theorem 1]{yassaee2014} is indeed a corollary of Lemma \ref{propmeasure}.

Secondly, since the R\'enyi divergence $D_{1+s}(\cdot)$ is a non-decreasing $s$ and thus for all $s\geq 0$,
\begin{align}
D(P_{M_{\calT}E^n}\|\prod_{t\in\calT}U_{\calM_t}\times P_E^n)
&=C_1(M_{\calT}|E^n)\\*
&\leq C_{1+s}(M_{\calT}|E^n)\label{basis4proof}.
\end{align}
Thus, our results in Lemma \ref{propmeasure} can be used to analyze the secrecy constraints in key generation problems if the constraints are expressed in terms of KL divergences or mutual information terms as in existing literature. 

It is not apparent how one can use Lemma \ref{propmeasure} for this purpose. To illustrate this, in the following, we briefly discuss the case in a private key generation problem involving three terminals: two legitimate users Alice and Bob, observing sequences $A_1^n$ and $A_2^n$ respectively, and one illegitimate user Eve who has access to side information $A^n$. Let $\bF$ denote the public communication between Alice and Bob and let $K$ denote the secret key generated by them. To make sure the generated key is secure, we need $I(K;A^n,\bF)$ to be small and to make sure the generated key is uniform, we need to make $D(P_K\|\rmU_K)$ to be small where $\rmU_K$ is the uniform distribution over the alphabet of the secret key.

Note that in key generation problems, in the achievability part, the public communication  $\bF$ is usually the random binning $(M_1,M_2)$ of observations $(A_1^n,A_2^n)$ at legitimate users and the secret key $K$ is usually obtained by applying a hash function on a commonly agreed binning sequence (which is correlated with $(A_1^n,A_2^n)$). Thus, we have that
\begin{align}
D(P_{K A^n\bF}\|P_{\rmU}\times P_{A^n}P_{\bF})
&=D(P_K\|\rmU_K)+I(K;A^n,M_1,M_2)\\
&\leq D(P_{K M_1M_2A^n}\|\rmU_{K}\times\rmU_{M_1}\rmU_{M_2}\times P_A^n),
\end{align}
where $U_{M_i},~i\in[2]$ is the uniform distribution over the alphabet of random binning.

Therefore, using Lemma \ref{propmeasure} and \eqref{basis4proof}, we have
\begin{itemize}
\item if \eqref{secrecycondition} is satisfied, then the secrecy constraint satisfies $\frac{1}{n}D(P_{K A^n\bF}\|\rmU_K\times P_{A^n \bF})$ vanishes to zero as $n\to\infty$ and thus ensures \emph{weak secrecy}; 
\item if the right hand side of \eqref{strongsecrecy} is always positive, then the secrecy constraint $D(P_{K A^n\bF}\|\rmU_K\times P_{A^n \bF})$ decays exponentially fast and thus ensures \emph{strong secrecy}.
\end{itemize}

In the remaining of this paper, to illustrate in detail how the result in Lemma \ref{propmeasure} can be used in analyses of secrecy constraints, we consider a multiple private key generation problem and derive the capacity region for the problem under mild conditions.

\section{Private Key Capacity Region for CMS} 
\label{sec:capacityresult}
\subsection{Multiple Private Key Generation with a Helper}
\label{sec:4users}

Let $P_{A_0A_1A_2A_3}$ be a joint probability density function (pdf) of random variables $(A_0,A_1,A_2,A_3)$ defined on a continuous alphabet $\calA_0\times\calA_1\times\calA_2\times\calA_3$. We assume that the pdf $P_{A_0A_1A_2A_3}$ satisfies that for any non-empty set $\calS\subseteq\{0,1,2,3\}$, the (joint) differential entropy $h(A_{\calS})$ is finite, i.e.,
\begin{align}
|h(A_{\calS})|=|h(\{A_t\}_{t\in\calS})|<\infty.
\end{align}
Let $(A_0^n,A_1^n,A_2^n,A_3^n)$ be a sequence of continuous random variables generated i.i.d. according to a pdf $P_{A_0A_1A_2A_3}$.

\begin{figure}[t]
\centering
\setlength{\unitlength}{0.5cm}
\scalebox{0.8}{
\begin{picture}(20,9)
\linethickness{1pt}
\put(1.5,10.5){\makebox(0,0){$A_1^n$}}
\put(1.5,10){\vector(0,-1){1.5}}
\put(6.5,10.5){\makebox(0,0){$A_0^n$}}
\put(6.5,10){\vector(0,-1){1.5}}
\put(11.5,10.5){\makebox(0,0){$A_2^n$}}
\put(11.5,10){\vector(0,-1){1.5}}
\put(17.5,10.5){\makebox(0,0){$A_3^n$}}
\put(17.5,10){\vector(0,-1){1.5}}
\put(1,7){\framebox(3,1.5){Alice}}
\put(6,7){\framebox(3,1.5){Bob}}
\put(11,7){\framebox(3,1.5){Charlie}}
\put(16,7){\framebox(3,1.5){Helen}}
\put(2.5,7){\vector(0,-1){1.5}}
\put(2.5,5.5){\vector(0,1){1.5}}
\put(7.5,7){\vector(0,-1){1.5}}
\put(7.5,5.5){\vector(0,1){1.5}}
\put(12.5,7){\vector(0,-1){1.5}}
\put(12.5,5.5){\vector(0,1){1.5}}
\put(17.5,7){\vector(0,-1){1.5}}
\put(17.5,5.5){\vector(0,1){1.5}}
\put(3.5,8.5){\vector(0,1){1.5}}
\put(3.5,10.5){\makebox(0,0){$K_\rmA$}}
\put(8.5,8.5){\vector(0,1){1.5}}
\put(9,10.5){\makebox(0,0){$(K_{\rm{BA}},K_{\rm{BC}})$}}
\put(13.5,8.5){\vector(0,1){1.5}}
\put(13.5,10.5){\makebox(0,0){$K_\rmC$}}
\put(1,4){\framebox(18,1.5){Public Discussion}}
\put(10,4){\vector(0,-1){1.5}}
\put(8.5,1){\framebox(3,1.5){Eve}}

\end{picture}}
\caption{Multiple Private key Generation with a Helper~\cite{zhang2017}.}
\label{systemmodel}
\end{figure}
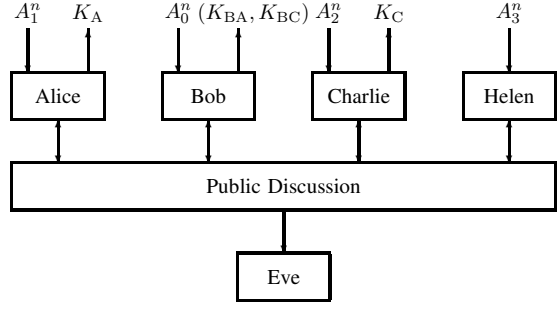

In this subsection, we revisit the multiple key generation model~\cite{zhang2017} by studied Zhang \emph{et al.} as shown in Figure \ref{systemmodel}.  In this model, there are four terminals: Alice has access to $A_1^n$, Bob has access to $A_0^n$, Charlie has access to $A_2^n$ and Helen has access to $A_3^n$. It is assumed that there is a noiseless public channel and all terminals talk interactively in $r$ rounds. Let the overall messages transmitted over the public channel be $\bF:=(F_1,\ldots,F_{4r})$. For $j=\{1,\ldots,4r\}$, $F_j$ is a function of $A_t^n$ and previous messages $F^{j-1}$ where $t=j\mod 4$. 

Let the alphabet of secret keys be $\calK_t:=\{1,\ldots,K_t\}$ for $t\in[2]$. Using the public messages $\bF$ and the source sequence $A_1^n$, Alice generates a private key $K_{\rmA}\in\calK_1$. Using $(\bF,A_0^n)$, Bob generates private keys $(K_{\rm{BA}},K_{\rm{BC}})\in\calK_1\times\calK_2$. Using $(\bF,A_2^n)$, Charlie generates $K_{\rmC}\in\calK_2$. We require that Alice and Bob agree on a private private key while Charlie and Bob agree on another private key, i.e. $K_\rmA=K_{\rm{BA}}$ and $K_\rmC=K_{\rm{BC}}$. A private key generation protocol consists of the public communication $\bF$. Note that in the above model, Helen is an untrusted helper who helps other terminals by transmitting messages over the public channel so that other terminals can obtain common sequences for subsequent key generations.

In \cite{zhang2017}, the authors considered four models with different secrecy requirements for \emph{discrete} memoryless sources, depending on whether $(K_\rmA,K_\rmC)$ is known by Helen and whether $K_\rmC$ is known by Alice. Our setting \emph{differs} from \cite{zhang2017} in the following two aspects:
\begin{enumerate}
\item We consider \emph{different secrecy requirements} on generated keys. To be specific, we require the private key $K_\rmA$ is only known by Alice and Bob and the private key $K_\rmC$ is only known by Bob and Charlie.
\item We consider \emph{continuous} memoryless sources, which requires different techniques in the analyses and derivation of fundamental limits concerning the performance of optimal protocols.
\end{enumerate}

We then give a formal definition of the capacity region of multiple private key generation with a helper, which concerns the asymptotic fundamental limits of optimal protocols.
\begin{definition}
\label{def:exponentregion}
A pair $(R_1,R_2)$ is said to be an achievable private key rate pair if there exists a sequence of private key generation protocols such that
\begin{align}
\lim_{n\to\infty}\max\big\{\Pr\{K_{\rmA}\neq K_{\rm{BA}}\},\Pr\{K_{\rmC}\neq K_{\rm{BC}}\}\big\}&=0\label{keymprivateeror},\\
\lim_{n\to\infty}D(P_{K_\rmA A_2^nA_3^n\bF}\|U_{\calK_1}\times P_{A_2^nA_3^n\bF})&=0\label{secrecymprivate1},\\
\lim_{n\to\infty} D(P_{K_\rmC A_1^nA_3^n\bF}\|U_{\calK_2}\times P_{A_1^nA_3^n\bF})&=0,\label{secrecymprivate2}\\
\liminf_{n\to\infty}\frac{1}{n}H(K_\rmA)\geq R_1,~\liminf_{n\to\infty}\frac{1}{n}H(K_\rmC)&\geq R_2\label{rateprivate}.
\end{align}
The closure of all achievable private key rate pairs is called the private key capacity region and denoted as $\calC_{\rm{MP}}$. 
\end{definition}
Note that \eqref{secrecymprivate1}, \eqref{secrecymprivate2} imply that i) the generated key $K_\rmA$ is almost uniform over $\calK_1$ and independent of $(\bF,A_2^n,A_3^n)$ and ii) $K_\rmC$ is almost uniform over $\calK_2$ and independent of $(\bF,A_1^n,A_3^n)$. Furthermore, the secrecy requirements in \eqref{secrecymprivate1}, \eqref{secrecymprivate2} are \emph{strong} in contrast to the weak ones in \cite{zhang2017}.

To present our result, we need the following definition. Let $\calR^*$ be the set of pairs $(R_1,R_2)$ such that
\begin{align}
R_1&\leq I(A_1;A_0|A_2,A_3),\\
R_2&\leq I(A_2;A_0|A_1,A_3),\\
R_1+R_2&\leq I(A_1,A_2;A_0|A_3)-I(A_1;A_2|A_3),
\end{align}
where the mutual information is calculated with respect to the pdf $P_{A_0A_1A_2A_3}$ or its induced pdfs.

\begin{theorem}
\label{strongcapacity}
The secrecy capacity region with an untrusted helper satisfies that
\begin{align}
\calR^*\subseteq\calC_{\rm{MP}}.
\end{align}
\end{theorem}
The proof of Theorem \ref{strongcapacity} is given in Section \ref{proofstrongc}. In the achievability proof, we first quantize the continuous source sequence similarly as in the proof of \cite[Theorem 1]{narayan2012}. Then, the terminals communicate over the public channel so that the quantized version of $(A_0^n,A_1^n)$ are decoded almost surely by Bob who observes $A_0^n$. The reliability analysis (cf. \eqref{keymprivateeror}) for key agreement proceeds similarly as the error exponent analysis for Slepian-Wolf coding introduced in \cite{gallager}. The secrecy analysis (cf. \eqref{secrecymprivate1}, \eqref{secrecymprivate1}) follows by invoking \eqref{strongsecrecy} in Lemma \ref{propmeasure}. Subsequently, we need to apply Fourier Motzkin Elimination to obtain the conditions on achievable rate pairs. Finally, as the quantization level goes to infinity, we show that any rate pair inside $\calR^*$ is achievable by exploring the relationship between the differential entropy of continuous random variables and the discrete entropy of the quantized random variables. 

We remark that Theorem \ref{strongcapacity} holds also for DMS, as can be gleaned from the proof. Furthermore, we can derive the achievable reliability-secrecy exponent which is positive for rate pairs inside $\calR^*$. We remark that Lemma \ref{propmeasure}, especially Eq. \eqref{strongsecrecy}, is critical to derive secrecy exponents~\cite{hayashi2011,chou2015} for key generation problems of DMS. This means that, we can not only show that Eq. \eqref{secrecymprivate1} and Eq. \eqref{secrecymprivate2} hold, but also derive a lower bound on the speed at which the secrecy constraints in Eq. \eqref{secrecymprivate1} and Eq. \eqref{secrecymprivate2} vanish to zero \emph{exponentially} as the length of observed sequences tends to infinity. This is yet another advantage of our method beyond quantization based techniques in \cite{chou2014separation} which could only be used to show that secrecy constraints vanish to zero \emph{but not the manner or the rate of decay}.

\begin{corollary}
\label{tightwhenmarkov}
For any pdf $P_{A_0A_1A_2A_3}$ such that the Markov chain $A_1-A_3-A_2$ holds, we have that $\calC_{\rm{MP}}=\calR^*$.
\end{corollary}
The proof of Corollary \ref{tightwhenmarkov} is given in Section \ref{prooftight}. When the Markov chain $A_1-A_3-A_2$ holds, we have $I(A_1;A_2|A_3)=0$. The achievability part follows from Theorem \ref{strongcapacity} and the converse part follows by judiciously adapting the converse techniques in \cite{narayan2012} to our setting. We remark that the proof techniques used to prove Theorem \ref{strongcapacity} and Corollary \ref{tightwhenmarkov} can also be applied to all the four models in \cite{zhang2017} and thus show that the capacity results in \cite{zhang2017} also hold for CMS with \emph{strong secrecy}.

\subsection{Generalization to a Cellular Model}
Recall that $\calT=\{1,\ldots,T\}$. For each $t\in\calT$, define an alphabet of keys as $\calK_t:=\{1,\ldots,K_t\}$. Let $(A_{\calT},A_0)$ be distributed according to a joint pdf $P_{A_0A_{\calT}}$ with zero mean vector and covariance matrix $\Sigma$. In this subsection, we consider a cellular model where there is a base station and $T$ terminals. This model is a generalization of our setting in Section \ref{sec:4users} in the spirit of \cite{csiszar2004secrecy} and has potential applications in internet of things where multiple terminals need to generate private keys with the help of other (potentially untrusted) terminals.

The base station observes the source sequence $A_0^n$ and for $t\in\calT$, terminal $t$ observes the source sequence $A_t^n$. Fix arbitrary subset $\calS$ of $\calT$. For each $t\in\calS$, terminal $t$ aims to generate a private key with the base station, concealed from all other terminals. We assume that the public communication is done in $r$ rounds over a noiseless public channel which is accessed by all terminals. Let $\bF$ denote the overall communication over the public channel. For each $t\in\calS$, given $\bF$ and $A_t^n$, terminal $t$ generates a private key $K_{\rmP,t}\in\calK_t$. Furthermore, given $A_0^n$ and $\bF$, the base station generates a sequence of private keys $\{K_{\rmB,t}\}_{t\in\calS}$. The goal of a good protocol is to enable the base station and each terminal $t\in\calS$ to generate an agreed private key, which is concealed from all other terminals.

The capacity region for this cellular model is defined as follows.
\begin{definition}
A tuple $R_{\calS}=\{R_t\}_{t\in\calS}$ is said to an achievable private key rate tuple if there exists a sequence of private key generation protocols ($\bF$) such that for each $t\in\calS$,
\begin{align}
\lim_{n\to\infty}\Pr\{K_{\rmP,t}\neq K_{\rmB,t}\}&=0\label{cellkeyerror},\\
\lim_{n\to\infty} D(P_{K_{\rmP,t}\bF A^n_{\calT\bigcap\{t\}^\rmc}}\|U_{\calK_t}\times P_{\bF A^n_{\calT\bigcap\{t\}^\rmc}})&=0\label{cellsecrecy},\\
\liminf_{n\to\infty}\frac{1}{n}H(K_{\rmP,t})&\geq R_t.
\end{align}
The closure of the set of all achievable private key rate tuples is called the private key capacity region and denoted as $\calC_{\rm{CP}}$.
\end{definition}

Before presenting the main results, we need the following definitions. Note that for $\calU\subseteq\calS$, $\calU\bigcup\calU^\rmc=\calT$.
\begin{align}
\calR_{\rm{in}}
&:=\Big\{R_{\calS}:~\forall~\emptyset\neq \calU\subseteq\calS:~\sum_{t\in\calU}R_t\leq \sum_{t\in\calU}h(A_t|A_{\calT\bigcap\{t\}^\rmc})-h(A_{\calU}|A_{\calU^\rmc}, A_0)\Big\}\\
\calR_{\rm{out}}&:=\Big\{R_{\calS}:~\forall~\emptyset\neq \calU\subseteq\calS:~\sum_{t\in\calU}R_t\leq h(A_{\calU}|A_{\calU^\rmc})-h(A_{\calU}|A_{\calU^\rmc}, A_0)\Big\}.
\end{align}
\begin{theorem}
\label{cellularcap}
The private key capacity region in the Cellular model satisfies that
\begin{align}
\calR_{\rm{in}}\subseteq\calC_{\rm{CP}}\subseteq\calR_{\rm{out}}.
\end{align}
\end{theorem}

The proof of Theorem \ref{cellularcap} is omitted since it is a generalization of the proofs of Theorem \ref{strongcapacity} and Corollary \ref{tightwhenmarkov}. In fact, we can recover the result in \ref{strongcapacity} and Corollary \ref{tightwhenmarkov} by letting $\calT=\{1,2,3\}$ and $\calS=\{1,2\}$. 

Here we provide only the proof sketch. In the achievability proof, we need to first quantize the source sequence at each terminal $t\in\calT$ and thus obtain $B_t^n$. Then, for $t\in\calS$, terminal $t$ sends a message $M_t\in\calM_t$ over the public channel and generates a private key $K_{\rmP,t}$ using $B_t^n$. For $t\in\calS^\rmc$, terminal $t$ sends the complete quantized source sequence. Thus, the public message $\bF=(M_\calS,B_{\calS^\rmc}^n)$. Given $A_0^n$ and $\bF$, the base station estimates $B_{\calS}^n$ and generate private key $K_{\rmB,t}$ using $\bF$ and the estimated sequences $\hatB_t^n$ for all $t\in\calS$. The error probability in key agreement is derived by using the distributed source coding idea and the secrecy analysis is done by invoking \eqref{strongsecrecy} in Lemma \ref{propmeasure}. Let $\tilR_t$ be the rate of the message at terminal $t$ and let the quantization interval go to zero. To satisfy \eqref{cellkeyerror} and \eqref{cellsecrecy}, we conclude that the rates should satisfy that for any positive $\delta$,
\begin{align}
R_t+\tilR_t&\leq H(A_t|A_{\calT\bigcap\{t\}^\rmc})-\delta,\\
\sum_{j\in\calU}\tilR_j&\geq H(A_\calU|A_{\calU^\rmc},A_0)+\delta.
\end{align}
for each $t\in\calS$ and for each non-empty subset $\calU$ of $\calS$. Without loss of generality, we can assume that $\calS=\{1,\ldots,|\calS|\}$. By applying the Fourier Motzkin Elimination successively to eliminate $\tilR_t$ for all $t\in\calS$, we conclude that any $R_{\calS}\in\calR_{\rm{in}}$ is achievable. 

The converse proof proceeds similarly as Corollary \ref{tightwhenmarkov} by assuming that there exists a super terminal observing $A_{\calU}^n$ and generating secret keys $K_{\rmP,\calU}:=\{K_{\rmP,t}\}_{t\in\calU}$ for each non-empty subset $\calU$ of $\calS$. This is possible since $T$ is finite and \eqref{cellsecrecy} implies that for any non-empty subset $\calU$ of $\calS$, we have that for any positive $\delta$ and sufficiently large $n$,
\begin{align}
D(P_{K_{\rmP,\calU}\bF A_{\calU^\rmc}^n}\|\prod_{t\in\calU}U_{\calM_t}\times P_{\bF A_{\calU^\rmc}^n})
&\leq \sum_{t\in\calU}\sum_{t\in\calU}D(P_{K_{\rmP,t}\bF A^n_{\calT\bigcap\{t\}^\rmc}}\|U_{\calK_t}\times P_{\bF A^n_{\calT\bigcap\{t\}^\rmc}})\\*
&\leq |\calU|\delta\leq T\delta.
\end{align}

\section{Proof of Theorem \ref{strongcapacity}}
\label{proofstrongc}
Throughout this section, we set $\calT=\{1,2,3\}$. 

\subsection{Coding Strategy}
Fix an integer $q$. Let $g_q:\calR\to[0,1,\ldots,2q^2]$ be a quantization function with quantization level $\Delta=\frac{1}{q}$ such that $g_q(a)=0$ if $a\leq -q$ or $a>q$ and $g_q(a)=\lceil q(q+a)\rceil$ if $a\in(-q,q]$. Note that the quantized random variable has a finite alphabet $\calB=[0,1,\ldots,2q^2]$ with the size $2q^2+1$. Applying the quantization function $Q$ on all $\{A_t\}_{t\in\calT}$ to obtain quantized version $\{B_t\}_{t\in\calT}$.  We first quantize the sequences $A_t^n$ using the function $g_q$ and obtain corresponding quantized sequences $B_t^n=g_q(A_t^n)$ for $t=0,1,2,3$.

Let $X^5=(X_1,X_2,X_3,X_4,X_5)$ be a sequence of independent random variables. Let $f_{X_t}$ be an universal$_2$ random hash function mapping from $\calB_t^n$ to $\calM_t=\{1,\ldots,N_t\}$ for $t=1,2,3$ where $X_t$ describes the stochastic behavior of the hash function. Similarly, let $f_{X_{t+3}}$ be random hash function mapping from $\calB_t^n$ to $\calK_t=\{1,\ldots,K_t\}$ for $t=1,2$. Furthermore, for any positive $\delta$, let $\log N_t=n\tilR_t$ for $t=1,2,3$ and $\log K_t=nR_t$ for $t=1,2$.

Codebook Generation:
The code book generated by Alice is $\calC_\rmA:=\bigcup_{a_1^n\calA_1^n}(f_{X_1}(g_q(a_1^n)),f_{X_4}(g_q(a_1^n)))$. The codebook generated by Charlie is $\calC_\rmC:=\bigcup_{a_2^n\calA_2^n}(f_{X_2}(g_q(a_2^n)),f_{X_5}(g_q(a_2^n)))$. The codebook generated by Helen is $\calC_\rmH:=\bigcup_{a_3^n\calA_3^n}f_{X_3}(g_q(a_3^n))$. The random codebook $\calC_{X^5}:=\{\calC_\rmA,\calC_\rmC,\calC_\rmH\}$ controlled by random variables $X^5$ is assumed to be known by all users Alice, Bob, Charlie and Helen.

Encoding: Recall that $B_t^n=g_q(A_t^n)$ for $t=0,1,2,3$. Given $A_1^n$, Alice sends $m_1:=f_{X_1}(B_1^n)$ over the public channel and takes $f_{X_4}(B_1^n)$ as the private key $K_\rmA$. Given $A_2^n$, Charlie sends $m_2:=f_{X_2}(B_2^n)$ and takes $f_{X_5}(B_2^n)$ as the private key $K_\rmC$. Given $A_3^n$, Helen sends $m_3:=f_{X_3}(B_3^n)$ over the public channel.

Decoding: Let $P_{B_0B_1B_1B_3}$ be induced by $P_{A_0A_1A_2A_3}$ and the quantization function $g_q$. Given the messages $\bF=(m_1,m_2,m_3)$ transmitted over the public discussion channel and the source sequence $A_0^n$, Bob uses maximum likelihood decoding to obtain $(\hatB_1^n,\hatB_2^n,\hatB_3^n)$, i.e.,
\begin{align}
(\hatB_1^n,\hatB_2^n,\hatB_3^n)
&:=\argmax_{\substack{(\tilb_1^n,\tilb_2^n,\tilb_3^n):\\f_{X_t}(\tilb_t^n)=m_t,~t=1,2,3}}P_{B_1B_2B_3|B_0}^n(\tilb_1^n,\tilb_2^n,\tilb_3^n|B_0^n).
\end{align}
Then, Bob claims that $K_{\rm{BA}}=f_{X_4}(\hatB_1^n)$ and $K_{\rm{BC}}=f_{X_5}(\hatB_3^n)$. 

\subsection{Analysis of Error Probability in Key Agreement}
Given the above coding strategy, we obtain that
\begin{align}
\max\big\{\Pr\{K_{\rmA}\neq K_{\rm{BA}}\},\Pr\{K_{\rmC}\neq K_{\rm{BC}}\}\big\}&\leq \Pr\Big\{(\hatB_1^n,\hatB_2^n,\hatB_3^n)\neq (B_1^n,B_2^n,B_3^n)\Big\}\label{reliable1}.
\end{align}
Note that the average is not only over all possible realizations of source sequences but also over all possible random universal$_2$ hash functions. Recall that in this section $\calT=\{1,2,3\}$ and all the quantized random variable have the same alphabet $\calB$. Given $\emptyset\neq\calS\subseteq\calT$ and $a_0^n\in\calR^n$, define the error events:
\begin{align}
\nn\calE_{\calS}&:=\Big\{\tilb_{\calT}^n\in\calB^{3n}:~\forall~t\in\calT,~f_{X_t}(\tilb_t^n)=m_t,~\forall~t\in\calS,~\tilb_t^n\neq B_t^n,\\*
&\qquad\quad\forall~t\in\calT\bigcap\calS^{\rmc},~\tilb_t^n=B_t^n,~P_{B_\calT|B_0}^n(\tilb_\calT^n|B_0^n)\geq P_{B_\calT|B_0}^n(B_\calT^n|B_0^n)\Big\}.
\end{align} 

Then, similarly as \cite{gallager}, we have that for any $\emptyset\neq\calS\subseteq\calT$ and arbitrary $s\in[0,1]$,
\begin{align}
\nn&\Pr\big\{\exists \tilb_{\calT}^n\in\calE_{\calS}\big\}\\*
&=\mathbb{E}_{X_{\calS}}\Big[\sum_{b_{\calT}^n}P_{B_{\calT}}^n(b_{\calT}^n)\Pr\big\{\exists \tilb_{\calT}^n\in\calE_{\calS}|b_{\calT}^n\big\}\Big]\\
&\leq \sum_{b_0^n,b_{\calT}^n}P_{B_0,B_{\calT}}^n(b_0^n,b_{\calT}^n)\Bigg(\sum_{\substack{\tilb_{\calS}^n}}1\{P_{B_{\calT}|B_0}^n(\tilb_{\calS}^n,b_{\calT\bigcap\calS^{\rmc}}^n|b_0^n)\geq P_{B_{\calT}|B_0}^n(b_{\calT}^n|b_0^n)\}\mathbb{E}_{X_{\calS}}\Big[1\{f_{X_t}(\tilb_t^n)=f_{X_t}^n(b_t^n),~t\in\calS\}\Big]\Bigg)^{s}\\
&\leq \sum_{b_0^n,b_{\calT}^n}P_{B_0,B_{\calT}}^n(b_0^n,b_{\calT}^n)\Bigg(\sum_{\substack{\tilb_{\calS}^n}}\frac{P_{B_{\calS}|B_0B_{\calT\bigcap\calS^{\rmc}}}^n(\tilb_{\calS}^n|b_0^n,b_{\calT\bigcap\calS^{\rmc}}^n)}{P_{B_{\calS}|B_0B_{\calT\bigcap\calS^{\rmc}}}^n(b_{\calK}^n|b_0^nb_{\calT\bigcap\calS^{\rmc}}^n)}\times \prod_{t\in\calS}\frac{1}{M_t}\Bigg)^{s}\\
&\leq \exp\Big(-ns\Big(\sum_{t\in\calS}\tilR_t-H_{1+s}^{\uparrow}(B_{\calS}|B_0,B_{\calT\bigcap\calK^{\rmc}})\Big)\label{reliable2},
\end{align}
where \eqref{reliable2} follows from the definition in \eqref{def:condentropyG}. Thus, invoking \eqref{reliable1} and \eqref{reliable2}, we obtain that
\begin{align}
\nn&\max\big\{\Pr\{K_{\rmA}\neq K_{\rm{BA}}\},\Pr\{K_{\rmC}\neq K_{\rm{BC}}\}\big\}\\*
&\leq \sum_{\emptyset\neq\calS\subseteq\calT}\Pr\big\{\exists \tilb_{\calT}^n\in\calE_{\calS}\big\}\\
&\leq \sum_{\emptyset\neq\calS\subseteq\calT}\exp\Big\{-n\max_{s\in[0,1]}s\Big(\sum_{t\in\calS}\tilR_t-H_{1+s}^{\uparrow}(B_{\calS}|B_0,B_{\calS^{\rmc}})\Big)\Big\}\\
&\leq (2^T-1)\times \exp\Big\{-n\min_{\emptyset\neq\calS\subseteq\calT}\max_{s\in[0,1]}s\Big(\sum_{t\in\calS}\tilR_t-H_{1+s}^{\uparrow}(B_{\calS}|B_0,B_{\calS^{\rmc}})\Big)\Big\}\label{derivedreliableexp}.
\end{align}

\subsection{Analysis of Secrecy Requirement}
Recall that  $U_{\calM_t}$ is the uniform distribution over $\calM_t$ for $t=1,2,3$ and let $U_{\calK_t}$ be the uniform distribution over $\calK_t$ for $t=1,2$. In the following, for simplicity, we will use $M_t$ to denote $f_{X_t}(B_t^n)$ for $t=1,2,3$. Given the coding strategy, we have
\begin{align}
D(P_{K_\rmA A_2^nA_3^n\bF}\|U_{\calK_1}\times P_{A_2^nA_3^n\bF})
&=D(P_{K_\rmA}\|U_{\calK_1})+I(K_\rmA;A_2^n,A_3^n,M_3)\\
&=D(P_{K_\rmA}\|U_{\calK_1})+I(K_\rmA;M_1)+I(K_\rmA;A_2^n,A_3^n,M_2,M_3|M_1)\\
&\leq D(P_{K_\rmA}\|U_{\calK_1})+I(K_\rmA;M_1)+I(K_\rmA,M_1;A_2^n,A_3^n)\label{explain1}\\
&\leq D(P_{K_\rmA M_1A_2^nA_3^n}\|U_{\calK_1}\times P_{U_1}\times P_{A_2,A_3}^n)\label{explain2}.
\end{align}
where \eqref{explain1} holds because $M_2$ is a function of $A_2^n$ and $M_3$ is a function of $A_3^n$, thus
\begin{align}
I(K_\rmA;A_2^n,A_3^n,M_2,M_3|M_1)
&=I(K_\rmA,M_1;A_2^n,A_3^n,M_2,M_3)-I(M_1;A_2^n,A_3^n,M_2,M_3)\\
&=I(K_\rmA,M_1;A_2^n,A_3^n)-I(M_1;A_2^n,A_3^n)\\
&\leq I(K_\rmA,M_1;A_2^n,A_3^n);
\end{align}
\eqref{explain2} holds because 
\begin{align}
\nn&D(P_{K_\rmA M_1A_2^nA_3^n}\|U_{\calK_1}\times P_{U_1}\times P_{A_2,A_3}^n)\\*
&=D(P_{K_\rmA M_1A_2^nA_3^n}\|P_{K_\rmA M_1}\times P_{A_2,A_3}^n)+D(P_{K_\rmA M_1}\times P_{A_2,A_3}^n\|U_{\calK_1}\times P_{U_1}\times P_{A_2,A_3}^n)\\
&=I(K_\rmA,M_1;A_2^n,A_3^n)+I(K_\rmA;M_1)+D(P_{K_\rmA}\|U_{\calK_1})+D(P_{M_1}\|P_{U_1})\\
&\geq I(K_\rmA,M_1;A_2^n,A_3^n)+I(K_\rmA;M_1)+D(P_{K_\rmA}\|U_{\calK_1}).
\end{align}

Using the result in \eqref{explain2} and invoking \eqref{strongsecrecy} in Lemma \ref{propmeasure} by replacing $E$ with $(A_2,A_3)$, we obtain that
\begin{align}
\liminf_{n\to\infty}-\frac{1}{n}\log\mathbb{E}_{X^5}\Big[D(P_{K_\rmA A_2^nA_3^n\bF}\|U_{\calK_1}\times P_{A_2^nA_3^n\bF})
\Big]\geq \max_{\theta\in[0,1]}\theta\Big(H_{1+\theta}(B_1|A_2,A_3)-R_1-\tilR_1\Big)\label{secrecypart1}.
\end{align}
Similarly as \eqref{secrecypart1}, we have
\begin{align}
\liminf_{n\to\infty}-\frac{1}{n}\log\mathbb{E}_{X^5}\Big[D(P_{K_\rmC A_1^nA_3^n\bF}\|U_{\calK_2}\times P_{A_1^nA_3^n\bF})
\Big]\geq \max_{\theta\in[0,1]}\theta\Big(H_{1+\theta}(B_2|A_1,A_3)-R_2-\tilR_2\Big)\label{secrecypart2}.
\end{align}

\subsection{Analysis of Capacity Region}
\begin{lemma}
\label{positiveexp}
Using the results in \eqref{derivedreliableexp}, \eqref{secrecypart1} and \eqref{secrecypart2}, we conclude that if $(R_1,R_2,\tilR_1,\tilR_2,\tilR_3)$ satisfies that for any positive $\delta$,
\begin{align}
\sum_{t\in\calS}\tilR_t&\geq H(B_{\calS}|B_{\calS^\rmc},B_0)+\delta,~\forall~\emptyset\neq\calS\subseteq\calT\label{reliareq},\\
R_1+\tilR_1&\leq H(B_1|A_2,A_3)-\delta\label{sreq1},\\
R_2+\tilR_2&\leq H(B_2|A_1,A_3)-\delta\label{sreq2},
\end{align}
then 
\begin{align}
\min_{\emptyset\neq\calS\subseteq\calT}\max_{s\in[0,1]}s\Big(\sum_{t\in\calS}\tilR_t-H_{1+s}^{\uparrow}(B_{\calS}|B_0,B_{\calS^\rmc})\Big)&>0\label{positiveexp1},\\
\max_{\theta\in[0,1]}\theta\Big(H_{1+\theta}(B_1|A_2,A_3)-R_1-\tilR_1\Big)&>0,\\
\max_{\theta\in[0,1]}\theta\Big(H_{1+\theta}(B_1|A_2,A_3)-R_1-\tilR_1\Big)&>0\label{positiveexp2},
\end{align}
\end{lemma}
The proof of Lemma \ref{positiveexp} follows from the properties of R\'enyi conditional entropy and thus omitted. By applying Fourier Motzkin Elimination to \eqref{reliareq} to \eqref{sreq2}, we obtain that $(R_1,R_2)$ should satisfy that 
\begin{align}
R_1&\leq H(B_1|A_2,A_3)-H(B_1|B_0,B_2,B_3)-2\delta,\\
R_2&\leq H(B_2|A_1,A_3)-H(B_2|B_0,B_1,B_3)-2\delta,\\
R_1+R_2&\leq H(B_1|A_2,A_3)+H(B_2|A_1,A_3)-H(B_1,B_2|B_0,B_3)-4\delta.
\end{align}
Recall that $\Delta=\frac{1}{q}$ is the quantization interval. Similarly as \cite[Lemma 3.1]{narayan2012} (see also \cite[Theorem 8.3.1]{cover2012elements}), we obtain the following result.
\begin{lemma}
\label{linkquantizetoGau}
\begin{align}
\nn&\lim_{\Delta\to 0} \Big(H(B_1|A_2,A_3)-H(B_1|B_0,B_2,B_3)\Big)\\*
&=h(A_1|A_2,A_3)-h(A_1|A_0,A_2,A_3)=I(A_0;A_1|A_2,A_3)\label{linkgau1},\\
\nn&\lim_{\Delta\to 0} \Big(H(B_2|A_1,A_3)-H(B_2|B_0,B_1,B_3)\Big)\\*
&=h(A_2|A_1,A_3)-h(A_2|A_0,A_1,A_3)=I(A_0;A_2|A_1,A_3)\label{linkgau2},\\
\nn&\lim_{\Delta\to 0}\Big(H(B_1|A_2,A_3)+H(B_2|A_1,A_3)-H(B_1,B_2|B_0,B_3)\Big)\\*
&=h(A_1|A_2,A_3)+h(A_2|A_1,A_3)-h(A_1,A_2|A_0,A_3)=I(A_0;A_1,A_2|A_3)-I(A_1;A_2|A_3)\label{linkgau3}.
\end{align}
\end{lemma}
The proof of Lemma \ref{linkquantizetoGau} is given in Appendix \ref{prooflinkquan}.

Invoking Lemma \ref{linkquantizetoGau} and letting $\delta\downarrow 0$, we have shown that average over all the random codebooks controlled by random variables $X^5$, if $(R_1,R_2)\in\calR^*$, then \eqref{keymprivateeror}, \eqref{secrecymprivate1} and \eqref{secrecymprivate2} are satisfied and thus $(R_1,R_2)$ is an achievable private key rate pair. The argument that there exists a deterministic codebook satisfying \eqref{keymprivateeror}, \eqref{secrecymprivate1} and \eqref{secrecymprivate2} can be done similarly as \cite{chou2015} and thus omitted.

\section{Proof of Corollary \ref{tightwhenmarkov}}
\label{prooftight}

\subsection{Preliminaries}
For $\calS\subseteq\calT$ and $\calW\subseteq\calA^\rmc$, let
\begin{align}
\calH(\calS|\calW)&:=\{\calU\subseteq\calW^\rmc:~\calU\neq\emptyset,~\calS\not\subseteq\calU\},~\label{def:calH}\\
\calH_i(\calS|\calW)&:=\{\calU\subseteq\calW^\rmc:~\calU\neq\emptyset,~\calS\not\subseteq\calU,~i\in\calU\},\label{def:calHi}\\
\Lambda(\calS|\calW)&:=\{\lambda:\sum_{\calU\in\calH_i(\calS|\calW)}\lambda_{\calU}=1,~\forall~i\in\calW^\rmc:~\calH_i(\calS|\calW)\neq \emptyset\}\label{def:Lambda}.
\end{align}
Similarly as \cite[Lemma 3.2]{narayan2012}, we can prove the following result.
\begin{lemma}
\label{converseaux}
Fix an integer $n$. Let $Z$ be a random variable jointly distributed with $A_{\calT}^n$.
\begin{enumerate}
\item For any $\lambda\in\Lambda(\calS|\calW)$, we have
\begin{align}
\!\!\!\!\!\!\!\! h(A_{\calT}^n|A_{\calW}^n,Z)-\sum_{\calU\in\calH(\calS|\calW)}\lambda_{\calU}h(A_{\calU}^n|h_{\calU^\rmc}^n,Z)&\geq 0\label{aux1};
\end{align}
\item For any $t\in\calW^\rmc$, let $V_t$ be a function of $(X_t,Z)$, then for any $\lambda\in\Lambda(\calS|\calW)$,
\begin{align}
\nn&h(A_{\calT}^n|A_{\calW}^n,Z)-\sum_{\calU\in\calH(\calS|\calW)}\lambda_{\calU}h(A_{\calU}^n|A_{\calU^\rmc}^n,Z)\\*
&=h(A_{\calT}^n|A_{\calW}^n,Z,V_t)-\sum_{\calU\in\calH(\calS|\calW)}\lambda_{\calU}h(A_{\calU}^n|A_{\calU^\rmc}^n,Z,V_t)+\sum_{\calU\in\calH_t(\calS|\calW)}I(V_t;A_{\calU^\rmc\bigcap\calW^\rmc}^n|A_{\calW}^n,Z)\label{aux2}.
\end{align}
\end{enumerate}
\end{lemma}

\subsection{Converse Proof}

Fix any secret key protocol with public message $\bF$ such that \eqref{keymprivateeror} to \eqref{rateprivate} are satisfied. We first consider keys $K_\rmA$ and $K_{\rm{BA}}$ only to derive an upper bound for $R_1$. Invoking \eqref{keymprivateeror} to \eqref{rateprivate}, we have that for sufficiently large $n$ and any positive $\delta$,
\begin{align}
\Pr\{K_\rmA\neq K_{\rm{BA}}\}&\leq \delta,\\
D(P_{K_\rmA A_2^nA_3^n\bF}\|U_{\calK_1}\times P_{A_2^nA_3^n\bF})&\leq \delta,\label{k1onlyreq}\\
\frac{1}{n}H(K_\rmA)&\geq R_1-\delta.
\end{align}

Recall that $\bF=(F_1,\ldots,F_{4r})$ are the total communication of $r$ and $F_j$ is a function of $A_t^n$ and $F^{j-1}$ where $t=j\mod 4$. Let $\bF_1:=\{F_j:~j\mod 4=0~\mathrm{or}~1\}$ and $\bF_2=\bF_1^\rmc$. Set $T=4$, $\calT=\{0,1,2,3\}$, $\calS=\{0,1\}$, $W=2$ and $\calW=\{2,3\}$. Thus, $\calH(\calS|\calW)=\{\{0\},\{1\}\}$. Invoking \eqref{def:calH} to \eqref{def:Lambda}, we obtain that
\begin{align}
h(A_0^n,A_1^n|A_2^n,A_3^n)-h(A_0^n|A_1^n,A_2^n,A_3^n)-h(A_1^n|A_0^n,A_2^n,A_3^n)
&=h(A_\calT^n|A_{\calW}^n)-\max_{\lambda\in\Lambda(\calS|\calW)}\sum_{\calU\in\calH(\calS|\calW)}\lambda_{\calU}h(A_{\calU}^n|A_{\calU^\rmc}^n).\end{align}
Invoking \eqref{aux2} with $Z=\emptyset$ and $V_0=F_0$ and noting that the summation of mutual information terms are non-negative, we obtain that
\begin{align}
h(A_{\calT}^n|A^n_{\calW})-\sum_{\calU\in\calH(\calS|\calW)}\lambda_{\calU}h(A^n_{\calU}|A^n_{\calU^\rmc})
&\geq h(A^n_{\calT}|A^n_{\calW},F_0)-\sum_{\calU\in\calH(\calS|\calW)}\lambda_{\calU}h(A^n_{\calU}|A^n_{\calU^\rmc},F_0)\\
&\geq h(A^n_{\calT}|A^n_{\calW},F_0,F_1)-\sum_{\calU\in\calH(\calS|\calW)}\lambda_{\calU}h(A^n_{\calU}|A^n_{\calU^\rmc},F_0,F_1)\label{successive1}\\
&\geq h(A^n_{\calT}|A^n_{\calW},\bF_1)-\sum_{\calU\in\calH(\calS|\calW)}\lambda_{\calU}h(A^n_{\calU}|A^n_{\calU^\rmc},\bF_1)\label{successivemore}\\
&=h(A^n_{\calT}|A^n_{\calW},\bF)-\sum_{\calU\in\calH(\calS|\calW)}\lambda_{\calU}h(A^n_{\calU}|A^n_{\calU^\rmc},\bF)\label{f2functaw},
\end{align}
where \eqref{successive1} follows by invoking \eqref{aux2} with $Z=F_0$ and $V_1=F_1$; \eqref{successivemore} follows by invoking \eqref{aux2} for $r(T-W)$ times successively; \eqref{f2functaw} follows because
\begin{align}
h(A_{\calT}^n|A_{\calW}^n,\bF_1)
&=h(A_{\calT}^n|A_{\calW}^n,\bF_1,F_{T-W})\label{functionof1}\\
&=h(A_{\calT}^n|A_{\calW}^n,\bF_1,F_{T-W},F_{T-W+1})\label{functionof2}\\
&=\ldots\\
&=h(A_{\calT}^n|A_{\calW}^n,\bF_1,\bF_2)\label{functionof},
\end{align}
and
\begin{align}
h(A_{\calU}^n|A_{\calU^\rmc}^n,\bF_1)=h(A_{\calU}^n|A_{\calU^\rmc}^n,\bF),
\end{align}
where \eqref{functionof1} follows since $F_{T-W}$ is a function of $A_{\calW}^n$ and $\bF_1$; \eqref{functionof2} follow since $F_{T-W+1}$ is a function of $(\bF_1,A_{\calW}^n,F_{T-W})$ and \eqref{functionof} follows by using the same idea successively for $WT$ times.

Note that $K_\rmA$ is a function of $A_1^n$ and $\bF$. Continuing from \eqref{f2functaw} and invoking \eqref{aux2} in Lemma \ref{converseaux} with $t=1$, $Z=\bF$, $V_1=K_\rmA$, we obtain that
\begin{align}
\nn&h(A^n_{\calT}|A^n_{\calW},\bF)-\sum_{\calU\in\calH(\calS|\calW)}\lambda_{\calU}h(A^n_{\calU}|A^n_{\calU^\rmc},\bF)\\*
&=h(A_{\calT}^n|A_{\calW}^n,\bF, K_\rmA)-\sum_{\calU\in\calH(\calS|\calW)}\lambda_{\calU}h(A_{\calU}^n|A_{\calU^\rmc}^n,\bF, K_\rmA)+\sum_{\calU\in\calH_1(\calS|\calW)}I(K_\rmA;A_{\calU^\rmc\bigcap\calW^\rmc}^n|A_{\calW}^n,\bF)\\
&\geq \sum_{\calU\in\calH_1(\calS|\calW)}I(K_\rmA;A_{\calU^\rmc\bigcap\calW^\rmc}^n|A_{\calW}^n,\bF)\label{useaux1}\\
&=I(K_\rmA;A_1^n|A_2^n,A_3^n,\bF)\label{usesetvalues}\\
&=h(K_\rmA|A_2^n,A_3^n,\bF)-h(K_\rmA|A_1^n,A_2^n,A_3^n,\bF)\\
&=h(K_\rmA)-I(K_\rmA;A_2^n,A_3^n,\bF)\label{krmafunc}\\
&\geq h(K_\rmA)-\delta\label{usereqk1}.
\end{align}
where \eqref{useaux1} follows from \eqref{aux1} in Lemma \ref{converseaux} by setting $Z=(\bF,K_\rmA)$; \eqref{usesetvalues} follows from the settings $\calT=\{0,1,2,3\}$, $\calS=\{0,1\}$, $\calW=\{2,3\}$, and $\calH_1(\calS|\calW)=\{\{1\}\}$; \eqref{krmafunc} follows since $K_\rmA$ is the function of $A_1^n$ and $\bF$; \eqref{usereqk1} follow by noting that $I(K_\rmA;A_2^nA_3^n\bF)\leq D(P_{K_\rmA A_2^nA_3^n\bF}\|U_{\calK_1}\times P_{A_2^nA_3^n\bF})\leq \delta$ and using \eqref{k1onlyreq}.

Therefore, invoking \eqref{f2functaw} and \eqref{usereqk1}, we conclude that
\begin{align}
R_1&\leq \liminf_{n\to\infty}\frac{1}{n}H(K_\rmA)\\
&\leq \liminf_{n\to\infty}\left(I(A_0;A_1|A_2,A_3)+\frac{\delta}{n}\right)\\
&=I(A_0;A_1|A_2,A_3)\label{conversepart1}.
\end{align}
Similarly as \eqref{conversepart1}, by considering the generation of $K_\rmC$ and $K_{\rm{BC}}$ only, we obtain that 
\begin{align}
R_2&\leq \liminf_{n\to\infty}\frac{1}{n}H(K_\rmC)\leq I(A_0;A_2|A_1,A_3).
\end{align}

Finally, we derive the bound on the sum rate. Invoking \eqref{keymprivateeror}, we obtain that for any $\delta$,
\begin{align}
\Pr\Big\{(K_\rmA,K_\rmC)\neq (K_{\rm{BA}},K_{\rm{BC}})\Big\}&\leq \max\big\{\Pr\{K_\rmA\neq K_{\rm{BA}}\},\Pr\{K_\rmC\neq K_{\rm{BC}}\}\big\}\\
&\leq 2\delta\label{newerror}.
\end{align}
Recall that $K_\rmA$ is a function of $(\bF,A_1^n)$ and $K_\rmC$ is a function of $(\bF,A_2^n)$. Invoking \eqref{secrecymprivate1} and \eqref{secrecymprivate2}, we obtain that
\begin{align}
2\delta
&\geq D(P_{K_\rmA A_2^nA_3^n\bF}\|U_{\calK_1}\times P_{A_2^nA_3^n\bF})+D(P_{K_\rmC A_1^nA_3^n\bF}\|U_{\calK_2}\times P_{A_1^nA_3^n\bF})\\
&=D(P_{K_\rmA}\|U_{\calK_1})+D(P_{K_\rmC}\|U_{\calK_2})+I(K_\rmA;A_2^n,A_3^n,\bF)+I(K_\rmC;A_1^n,A_3^n,\bF)\\
&=D(P_{K_\rmA}\|U_{\calK_1})+D(P_{K_\rmC}\|U_{\calK_2})+I(K_\rmA;K_\rmC,A_2^n,A_3^n,\bF)+I(K_\rmC;A_1^n,A_3^n,\bF)\\
&=D(P_{K_\rmA}\|U_{\calK_1})+D(P_{K_\rmC}\|U_{\calK_2})+I(K_\rmA;K_\rmC)+I(K_\rmA;A_2^n,A_3^n,\bF|K_\rmc)+I(K_\rmC;A_1^n,A_3^n,\bF)\\
&\geq D(P_{K_\rmA}\|U_{\calK_1})+D(P_{K_\rmC}\|U_{\calK_2})+I(K_\rmA;K_\rmC)+I(K_\rmA;A_3^n,\bF|K_\rmc)+I(K_\rmC;A_3^n,\bF)\\
&=D(P_{K_\rmA K_\rmC A_3^n\bF}\|U_{\calK_1}\times U_{\calK_2}\times P_{A_3^n\bF})\label{newsecrecy},
\end{align}
and 
\begin{align}
\delta&\geq D(P_{K_\rmA A_2^nA_3^n\bF}\|U_{\calK_1}\times P_{A_2^nA_3^n\bF})\\
&\geq I(K_\rmA;K_\rmC).
\end{align}
Thus, we have
\begin{align}
\liminf_{n\to\infty}\frac{1}{n}h(K_\rmA K_\rmC)
&=\liminf_{n\to\infty}\frac{1}{n}(h(K_\rmA)+h(K_\rmC)-I(K_\rmA;K_\rmC))\\
&\geq \liminf_{n\to\infty}\frac{1}{n}(h(K_\rmA)+h(K_\rmC)-\delta)\\
&\geq R_1+R_2.
\end{align}

Then, let us consider a super terminal observing $(A_1^n,A_2^n)$ and generate private keys $(K_\rmA,K_\rmC)$. With the requirement in \eqref{newerror}, \eqref{newsecrecy}, similarly as \eqref{conversepart1}, we conclude that
\begin{align}
R_1+R_2&\leq \liminf_{n\to\infty}\frac{1}{n}h(K_\rmA K_\rmC)
\leq I(A_0;A_1,A_2|A_3)+4\delta.
\end{align}
The proof of Corollary \ref{tightwhenmarkov} is now complete.

\section{Conclusion}
\label{sec:conc}
We first presented the output statistics of hash functions under the R\'enyi divergence criterion in Lemma \ref{propmeasure}. Lemma \ref{propmeasure} is a generalization of the result in \cite{hayashi2011} to the multi terminal case and the strict generalization of the output statistics in \cite[Theorem 1]{yassaee2014} where the variation distance is used as the security measure. Subsequently, we illustrated the power of Lemma \ref{propmeasure} in analyzing secrecy constraints by deriving the capacity region of the multiple private key generation problem with a helper for CMS. The converse proof follows by judiciously adapting the techniques in \cite{narayan2012} to the case with correlated side information at untrusted terminals. 

We then briefly discuss the future research directions. First, one can apply Lemma \ref{propmeasure} to analyze secrecy constraints for other key generation problems for CMS, such as the multi-terminal private key generation problem~\cite[Theorem 2]{csiszar2004secrecy} and the secret-private key generation problem with three terminals~\cite{ye2005isit}. Furthermore, as shown in Theorems \ref{strongcapacity}, \ref{cellularcap} and Corollary \ref{tightwhenmarkov}, the capacity region for multiple private key generation is not tight in general. One may nail down the exact capacity region. Second, one may derive second-order asymptotics for multi-terminal key generation problems and thus extend the results of \cite{hayashi2016ssecond}. In order to do so, for private key generation problems, one can potentially refer to \cite{chan2010,tyagi2015} to derive the converse part and extend the achievability scheme in \cite{hayashi2016ssecond} to the multi-terminal case. Note that in \cite{tyagi2015,hayashi2016ssecond}, the secrecy measure is the variational distance. Finally, one can explore the fundamental limits of the key generation problems with R\'enyi divergence as the security measure, as proposed in \cite{hayashi2017}. For capacity results, the achievability part can probably be done by using Lemma \ref{propmeasure} or extending the results in \cite{hayashi2017}.

\appendix
\subsection{Proof of Lemma \ref{propmeasure}}
\label{proofpropmeasure}
For simplicity, we consider $n=1$ and discrete variable $E$ (i.e., $\calE$ is finite). The case for continuous variable $E$ and for any $n\in\bbN$ can be done similarly by replacing the summation over $e\in\calE$ with corresponding integrals and using the i.i.d. nature of source sequences. For simplicity, we use $\calA_{\calT}$ to denote $\prod_{t\in\calT}\calA_t$, $\calM_{\calT}$ to denote $\prod_{t\in\calT}\calM_t$ and 
$1\{f_{X_{\calT}}(a_{\calT})=m_{\calT}\}$ to denote $\prod_{t\in\calT}1\{f_{X_t}(a_t)=m_t\}$ for all $a_\calT\in\calA_{\calT}$ and $m_\calT\in\calM_{\calT}$. Given $a_{\calT}\in\calA_{\calT}$, for any subset $\calS$ of $\calT$, define
\begin{align}
\calB_{\calS}:=\{\bara_{\calT}:~\bara_{\calS}=a_{\calS},~\mathrm{and}~\forall~t\notin\calS,~\bara_t\neq a_t\}.
\end{align}
Thus, we have
\begin{align}
\calA_{\calT}=\bigcup_{\emptyset\neq \calS\subseteq\calT}\calB_{\calS}.
\end{align}

\subsubsection{Proof of Claim (i)}
Fix $a_{\calT}$ and $e$. For any non-empty set $\calS\subseteq\calT$, we have that 
\begin{align}
\mathbb{E}_{X_{\calT}} \Big[\sum_{\bara_{\calT}\in\calB_{\calS}}1\{f_{X_{\calT}}(\bara_{\calT})=f_{X_\calT}(a_{\calT})\}P_{A_{\calT}|E}(\bara_{\calT}|e)\Big]&\leq P_{A_{\calS}|E}(a_{\calS}|e)\times \prod_{t\in(\calT-\calS)}\frac{\varepsilon}{N_t}\label{propepsilonofhash},
\end{align}
where \eqref{propepsilonofhash} follows from the $\varepsilon$-almost universal property of hash functions $f_{X_t}$ for all $t\in\calT$. Similarly, if $\calS=\emptyset$, then we have
\begin{align}
\mathbb{E}_{X_{\calT}} \Big[\sum_{\bara_{\calT}\in\calB_{\calS}}1\{f_{X_{\calT}}(\bara_{\calT})=f_{X_\calT}(a_{\calT})\}P_{A_{\calT}|E}(\bara_{\calT}|e)\Big]&\leq \prod_{t\in\calT}\frac{\varepsilon}{N_t}\label{propepsilonofhash2}.
\end{align}

Therefore, invoking \eqref{propepsilonofhash} and \eqref{propepsilonofhash2}, we obtain that
\begin{align}
\nn&\mathbb{E}_{X_{\calT}} \Big[\sum_{\bara_{\calT}}1\{f_{X_{\calT}}(\bara_{\calT})=f_{X_\calT}(a_{\calT})\}P_{A_{\calT}|E}(\bara_{\calT}|e)\Big]\\*
&=\mathbb{E}_{X_{\calT}} \Big[\sum_{\calS\subseteq\calT}\sum_{\bara_{\calT}\in\calB_{\calS}}1\{f_{X_{\calT}}(\bara_{\calT})=f_{X_\calT}(a_{\calT})\}P_{A_{\calT}|E}(\bara_{\calT}|e)\Big]\\
&=\sum_{\calS\subseteq\calT}\mathbb{E}_{X_{\calT}} \Big[\sum_{\bara_{\calT}\in\calB_{\calS}}1\{f_{X_{\calT}}(\bara_{\calT})=f_{X_\calT}(a_{\calT})\}P_{A_{\calT}|E}(\bara_{\calT}|e)\Big]\\
&\leq \sum_{\emptyset\neq\calS\subseteq\calT} P_{A_{\calS}|E}(a_{\calS}|e)\times \prod_{t\in(\calT-\calS)}\frac{\varepsilon}{N_t}+\prod_{t\in\calT}\frac{\varepsilon}{N_t}\label{tobeused}.
\end{align}

Thus, we obtain that
\begin{align}
\nn&\mathbb{E}_{X_{\calT}}\Big[\exp(-sH_{1+s}(M_{\calT}|E)\Big]\\*
&=\mathbb{E}_{X_{\calT}}
\Bigg[\sum_e P_E(e)\sum_{m_{\calT}\in\calM_{\calT}}\Big(\sum_{a_{\calT}}1\{f_{X_{\calT}}(a_{\calT})=m_{\calT}\}P_{A_{\calT}|E}(a_{\calT}|e)\Big)^{1+s}\Bigg]\\
&=\mathbb{E}_{X_{\calT}}
\Bigg[\sum_e P_E(e)\sum_{a_{\calT}}P_{A_{\calT}|E}(a_{\calT}|e)\Big(\sum_{\bara_{\calT}}1\{f_{X_{\calT}}(\bara_{\calT})=f_{X_\calT}(a_{\calT})\}P_{A_{\calT}|E}(\bara_{\calT}|e)\Big)^{s}\Bigg]\\
&\leq \sum_e P_E(e)\sum_{a_{\calT}}P_{A_{\calT}|E}(a_{\calT}|e)\Big(\mathbb{E}_{X_{\calT}} \Big[\sum_{\bara_{\calT}}1\{f_{X_{\calT}}(\bara_{\calT})=f_{X_\calT}(a_{\calT})\}P_{A_{\calT}|E}(\bara_{\calT}|e)\Big]\Big)^{s}\label{concavity}\\
&\leq \sum_e P_E(e)\sum_{a_{\calT}}P_{A_{\calT}|E}(a_{\calT}|e)\Big(\sum_{\emptyset\neq\calS\subseteq\calT} P_{A_{\calS}|E}(a_{\calS}|e)\times \prod_{t\in(\calT-\calS)}\frac{\varepsilon}{N_t}+\prod_{t\in\calT}\frac{\varepsilon}{N_t}\Big)^s\label{usehash}\\
&\leq \sum_e P_E(e)\sum_{a_{\calT}}P_{A_{\calT}|E}(a_{\calT}|e)\Big(\sum_{\emptyset\neq\calS\subseteq\calT} P_{A_{\calS}|E}^s(a_{\calS}|e)\prod_{t\in(\calT-\calS)}\frac{\varepsilon^s}{N_t^s}+\prod_{t\in\calT}\frac{\varepsilon^s}{N_t^s}\Big)\label{useinq}\\
&\leq \prod_{t\in\calT}\frac{\varepsilon^s}{N_t^s}+\sum_{\emptyset\neq\calS\subseteq\calT}\Big(\prod_{t\in(\calT-\calS)}\frac{\varepsilon^s}{N_t^s}\Big)\Big(\sum_e P_E(e)\sum_{a_{\calS}}P_{A_{\calS}|E}^{1+s}(a_{\calS}|e)\Big)\\
&=\prod_{t\in\calT}\frac{\varepsilon^s}{N_t^s}+\sum_{\emptyset\neq\calS\subseteq\calT}\Big(\prod_{t\in(\calT-\calS)}\frac{\varepsilon^s}{N_t^s}\Big)\exp(-sH_{1+s}(A_{\calS}|E))\label{achlast}.
\end{align}
where \eqref{concavity} follows from the concavity of the function $t^s$ for $s\in[0,1]$; \eqref{usehash} follows from \eqref{def:condryentropy} and \eqref{tobeused}; \eqref{useinq} follows from the inequality $(\sum_i a_i)^s\leq \sum_i a_i^s$ for $s\in[0,1]$~\cite[Problem 4.15(f)]{gallagerIT}; \eqref{achlast} follows from the definition in \eqref{def:condryentropy}.

Invoking \eqref{secrecym2} and \eqref{achlast}, we conclude that
\begin{align}
\nn&\mathbb{E}_{X_{\calT}}\Big[\exp(sC_{1+s}(M_{\calT}|E))\Big]\\*
&=\mathbb{E}_{X_{\calT}}\Big[\exp\Big(s\sum_{t\in\calT}\log N_t-sH_{1+s}(M_{\calT}|E)\Big)\Big]\\
&\leq \Big(\prod_{t\in\calT}N_t^s\Big)\times \Big(\prod_{t\in\calT}\frac{\varepsilon^s}{N_t^s}+\sum_{\emptyset\neq\calS\subseteq\calT}\Big(\prod_{t\in(\calT-\calS)}\frac{\varepsilon^s}{N_t^s}\Big)\exp(-sH_{1+s}(A_{\calS}|E))\Big)\\
&=\varepsilon^{sT}+\sum_{\emptyset\neq\calS\subseteq\calT}\varepsilon^{s(T-|\calS|)}\Big(\prod_{i\in\calS} N_t^s\Big) \exp(-sH_{1+s}(A_{\calS}|E))\label{achresult1}.
\end{align}
The proof of Claim (i) is thus completed.

\subsubsection{Proof of Claim (ii)}
Recall that 
\begin{align}
\log N_t=nR_t.
\end{align}

Given any $s\in(0,1]$ and any $\varepsilon\in\bbR_+$, we have
\begin{align}
\mathbb{E}_{X_{\calT}}\Big[C_{1+s}(M_{\calT}|E^n)\Big]&\leq \frac{1}{s}\log\left(\mathbb{E}_{X_{\calT}}\Big[\exp(sC_{1+s}(M_{\calT}|E))\Big]\right)\label{concavitylog}\\
&\leq \frac{1}{s}\log \Bigg(\varepsilon^{sT}+\sum_{\emptyset\neq\calS\subseteq\calT}\varepsilon^{s(T-|\calS|)}\exp\Big(-sn \big(H_{1+s}(A_{\calS}|E)-\sum_{t\in\calS}R_t\big)\Big)\Bigg)
\label{useclaimi}\\
&\leq |T\log\varepsilon|+\frac{1}{s}\log \Bigg(1+\sum_{\emptyset\neq\calS\subseteq\calT}\exp\Big(-sn \big(H_{1+s}(A_{\calS}|E)-\sum_{t\in\calS}R_t\big)\Big)\Bigg)\label{useeps<=1},
\end{align}
where \eqref{concavitylog} follows since $\exp(\mathbb{E}_{X_{\calT}}\big[sC_{1+s}(M_{\calT}|E)\big])\leq \mathbb{E}_{X_{\calT}}[\exp(sC_{1+s}(M_{\calT}|E))]$ due to the convexity of $\exp(z)$ in $z\in\bbR$; \eqref{useclaimi} follows from the result in \eqref{achresult1} and the fact that $(A_{\calS}^n,E^n)$ are a sequence of i.i.d. random variables, leading to $H_{1+s}(A_{\calS}^n|E^n)=nH_{1+s}(A_{\calS}|E)$; \eqref{useeps<=1} follows since i) $\log(z)$ is increasing in $z\in\bbR_+$ and ii) for any $\varepsilon\in\bbR_+$, with $g(\calS):=\exp\Big(-sn \big(H_{1+s}(A_{\calS}|E)-\sum_{t\in\calS}R_t\big)$, if $\varepsilon\in(0,1]$, then
\begin{align}
\varepsilon^{sT}+\sum_{\emptyset\neq\calS\subseteq\calT} \varepsilon^{s(T-|\calS|)}g(\calS)
&\leq 1+\sum_{\emptyset\neq\calS\subseteq\calT}g(\calS)
\end{align}
and if $\varepsilon>1$, then
\begin{align}
\varepsilon^{sT}+\sum_{\emptyset\neq\calS\subseteq\calT} \varepsilon^{s(T-|\calS|)}g(\calS)
&\leq \varepsilon^{sT}\times\Big(1+\sum_{\emptyset\neq\calS\subseteq\calT}g(\calS)\Big).
\end{align}

The proof of Claim (ii) is completed by invoking \eqref{useeps<=1}.

\subsection{Proof of Claim (iii)}

We then proceed to prove \eqref{strongsecrecy}. From now on, we take $\varepsilon=1$ and thus consider universal$_2$ hash functions. For any $s\in(0,1]$ and $\theta\in[s,1]$, using \eqref{achresult1}, we obtain that
\begin{align}
\mathbb{E}_{X_{\calT}}\Big[C_{1+s}(M_{\calT}|E^n)\Big]&\leq \mathbb{E}_{X_{\calT}}\Big[C_{1+\theta}(M_{\calT}|E^n)\Big]\\
&\leq \frac{1}{\theta}\log \Big(1+\sum_{\emptyset\neq\calS\subseteq\calT} \Big(\prod_{i\in\calS} M_i^r\theta\Big)\times \exp(-rH_{1+\theta}(A_{\calS}^n|E^n))\Big)\\
&\leq \frac{1}{r}\sum_{\emptyset\neq\calS\subseteq\calT} \exp\Big(-n\theta\big(H_{1+\theta}(A_{\calS}|E)-\sum_{t\in\calS}R_t\big)\Big)\\
&\leq \frac{2^T-1}{\theta}\times \max_{\emptyset\neq \calS\subseteq\calT}\exp\Big(-n\theta\big(H_{1+\theta}(A_{\calS}|E)-\sum_{t\in\calS}R_t\big)\Big)\\
&=\frac{2^T-1}{\theta}\times \exp\Big(-n\theta\min_{\emptyset\neq \calS\subseteq\calT}\big(H_{1+\theta}(A_{\calS}|E)-\sum_{t\in\calS}R_t\big)\Big)\label{bighelp2}.
\end{align}
Thus, invoking \eqref{bighelp2}, we obtain that for $s\in(0,1]$, we have
\begin{align}
\liminf_{n\to\infty}-\frac{1}{n}\log \mathbb{E}_{X_{\calT}}\Big[C_{1+s}(M_{\calT}|E^n)\Big]
&\geq \max_{\theta\in(s,1]}\min_{\emptyset\neq \calS\subseteq\calT}\big(H_{1+\theta}(A_{\calS}|E)-\sum_{t\in\calS}R_t\big).
\end{align}

Invoking \eqref{useeps<=1}, we obtain that $C_{1+s}(M_{\calT}|E^n)=O(n)$ Thus, recalling that $C_{1+s}(\cdot)$ is non-decreasing in $s$, we obtain that for $s=0$,
\begin{align}
\liminf_{n\to\infty}-\frac{1}{n}\log\mathbb{E}_{X_{\calT}}\Big[C_{1+s}(M_{\calT}|E^n)\Big]
&\geq \min\{0,\max_{\theta\in(0,1]}\theta\min_{\emptyset\neq \calS\subseteq\calT}\big(H_{1+\theta}(A_{\calS}|E)-\sum_{t\in\calS}R_t\big)\}\\
&=\max_{\theta\in[0,1]}\theta\min_{\emptyset\neq \calS\subseteq\calT}\big(H_{1+\theta}(A_{\calS}|E)-\sum_{t\in\calS}R_t\big).
\end{align}

\subsection{Proof of Lemma \ref{linkquantizetoGau}}
\label{prooflinkquan}
Here we only provide the proof of \eqref{linkgau1} since \eqref{linkgau2} and \eqref{linkgau3} can be proved similarly. Recall that for $t=0,1,2,3$, $B_t$ is the quantized version of $A_t$, i.e., $B_t=g_q(A_t)$. Define an auxiliary random variable $B_4:=\prod_{t=1}^3 1\{B_t\neq 0\}$. Then we have that
\begin{align}
H(B_1|A_2,A_3)
&=H(B_1,B_4|A_2,A_3)-H(B_4|B_1,A_2,A_3)\\
&=H(B_1,B_4|A_2,A_3)\label{B4func}\\
&=H(B_4|A_2,A_3)+H(B_1|A_2,A_3,B_4)\label{prooflinkstep1}.
\end{align}
where \eqref{B4func} follows since $B_4$ is function of $(B_1,B_2,B_3)$ and $B_t$ is a function of $A_t$ for $t=2,3$. Note that $(A_2,A_3)-(B_1,B_2,B_3)-B_4$ and $B_1-A_1-(A_2,A_3)-(B_2,B_3)$ form Markov chains. Hence, for any $(a_2,a_3,b_1,b_2,b_3)\in\calR^2\times\calB^3$,
\begin{align}
P_{A_2^3B_1^3B_4}(a_2,a_3,b_1^3,1)&=P_{A_2^3}(a_2^3)P_{B_2^3|A_2^3}(b_2^3|a_2^3)P_{B_1|A_2^3}(b_1|a_2^3)P_{B_4|B_1^3}(1|b_1^3)\\
&=P_{A_2^3}(a_2^3)\times\prod_{t=2}^31\{b_t=g_q(a_t)\}P_{B_1|A_2^3}(b_1|a_2^3)\times \prod_{t=1}^3 1\{b_t\neq 0\}
\end{align}
Thus, $P_{A_2^3 B_1^3B_4}(a_2^3,b_1^3,1)$ is non-zero if and only if $g_q(a_t)\neq 0$ for $t=1,2$ and $b_t=1$ for $t=1,2,3$. Let 
\begin{align}
\eta(a_2^3):=P_{B_4|A_2^3}(0|a_2^3).\label{def:eta}
\end{align}
Thus,
\begin{align}
\nn&H(B_1,B_2,B_3|A_2=a_2,A_3=a_3,B_4=1)\\*
&=-\sum_{b_1^3\in\{1,\ldots,2q^2\}^3}\frac{P_{A_2^3B_1^3B_4}(a_2^3,b_1^3,1)}{P_{A_2^3B_4}(a_2^3,1)}\log \frac{P_{A_2^3B_1^3B_4}(a_2^3,b_1^3,1)}{P_{A_2^3B_4}(a_2^3,1)}\\
&=-\sum_{b_1\in\{1,\ldots,2q^2\}}\frac{P_{B_1|A_2^3}(b_1|a_2^3)}{1-\eta(a_2^3)}\log\frac{P_{B_1|A_2^3}(b_1|a_2^3)}{1-\eta(a_2^3)}\\
&=\frac{\log(1-\eta(a_2^3))}{1-\eta(a_2^3)}-\frac{1}{1-\eta(a_2^3)}\sum_{b_1\in\{1,\ldots,2q^2\}}P_{A_1|A_2^3}(g_q^{-1}(b_1)|a_2^3)\Delta\log P_{A_1|A_2^3}(a_1|a_2^3)\Delta\label{meanvalue}\\
\nn&=\frac{\log(1-\eta(a_2^3))}{1-\eta(a_2^3)}-\frac{\log \Delta}{1-\eta(a_2^3)}\sum_{b_1\in\{1,\ldots,2q^2\}}P_{A_1|A_2^3}(g_q^{-1}(b_1)|a_2^3)\Delta\\*
&\qquad-\frac{1}{1-\eta(a_2^3)}\sum_{b_1\in\{1,\ldots,2q^2\}}P_{A_1|A_2^3}(g_q^{-1}(b_1)|a_2^3)\Delta\log P_{A_1|A_2^3}(a_1|a_2^3)\label{tobeused1},
\end{align}
where \eqref{meanvalue} follows from the mean value theorem, which states that for some $a_1$ such that $g_q(a_1)=b_1$,
\begin{align}
P_{B_1|A_2^3}(b_1|a_2^3)&=\int_{\substack{a_1:(b_1-1)\Delta-q<a_1\leq b_1\Delta-q}}P_{A_1|A_2^3}(a_1|a_2^3)\rmd a_1\\
&=P_{A_1|A_2^3}(a_1|a_2^3)\Delta.
\end{align} 

Let $\Sigma_1$ be the variance of $A_1$. Similarly as \cite[(66)-(67)]{narayan2012}, we obtain that as $\Delta=\frac{1}{q}\downarrow 0$, for any $(a_2,a_3)$
\begin{align}
\eta(a_2^3)
&=P_{B_4|A_2^3}(0|a_2^3)\\
&=\prod_{t=2}^3 1\{g(a_t)=0\}\times \Pr\{B_1=0\}\\
&=\prod_{t=2}^3 1\{g(a_t)=0\}\times \Pr\{A_1\notin[-q,q]\}\\
&\leq \exp\Big(-\frac{q^2}{2\Sigma_1}-\frac{1}{2}\log 2\pi\Sigma_1\Big)\to 0\label{etasmall}.
\end{align}
Let $h_b(x):=-x\log x-(1-x)\log(1-x)$ be the binary entropy function for $x\in[0,1]$. Invoking \eqref{etasmall}, we obtain that
\begin{align}
\lim_{\Delta\to 0}H(B_4|A_2,A_3)
&=\lim_{\Delta\to 0}\int_{a_2^3} P_{A_2^3}(a_2^3)H(B_4|a_2^3)\rmd_{a_2^3}\\
&\leq \lim_{\Delta\to 0}\int_{a_2^3} P_{A_2^3}(a_2^3)h_b(\eta(a_2^3))\rmd_{a_2^3}=0\label{link1part2}.
\end{align}

Invoking \eqref{tobeused1}, \eqref{etasmall} and noting that $B_t$ is a function of $A_t$ for $t=2,3$, we obtain that
\begin{align}
\lim_{\Delta\to0}H(B_1|A_2,A_3,B_4)
&=\lim_{\Delta\to0}H(B_1,B_2,B_3|A_2,A_3,B_4)\\
\nn&=\lim_{\Delta\to 0} \Big(
\int_{a_2^3}P_{A_2^3}(a_2^3)P_{B_4|A_2^3}(0|a_2^3)H(B_1|A_2=a_2,A_3=a_3,B_4=0)\rmd_{a_2^3}\\*
&\qquad+\int_{a_2^3}P_{A_2^3}(a_2^3)P_{B_4|A_2^3}(1|a_2^3) H(B_2|A_2=a_2,A_3=a_3,B_4=1)\rmd_{a_2^3}\Big)\\
&=h(A_1|A_2,A_3)\label{link1part3}.
\end{align}

Therefore, invoking \eqref{prooflinkstep1}, \eqref{link1part2}, \eqref{link1part3}, we obtain that
\begin{align}
\lim_{\Delta\to 0}H(B_1|A_2,A_3)=h(A_1|A_2,A_3).
\end{align}
The proof of \eqref{linkgau1} is complete if we show that
\begin{align}
\lim_{\Delta\to 0} H(B_1|B_0,B_2,B_3)=h(A_1|A_0,A_2,A_3)\label{link1toprove}.
\end{align}
For this purpose, define $B_5=\prod_{t=0}^31\{B_t\neq 0\}$ and $B_6:=\prod_{t\in\{0,2,3\}}\{B_t\neq 0\}$. Then, we have
\begin{align}
H(B_1|B_0,B_2,B_3)&=H(B_0^3)-H(B_0,B_2,B_3)\\
&=H(B_0^3,B_5)-H(B_0,B_2,B_3,B_6)\\
&=H(B_5)+H(B_0^3|B_5)-H(B_6)-H(B_0,B_2,B_3|B_6)\label{link1part4}.
\end{align}
Similarly as \cite[Equation (67)]{narayan2012}, we can show that for $t=5,6$,
\begin{align}
\lim_{\Delta\to 0} \Pr\{B_t=0\}=0.
\end{align}
Hence, we obtain that
\begin{align}
\lim_{\Delta\to 0}H(B_5)&=0\label{link1p5},\\
\lim_{\Delta\to 0}H(B_6)&=0\label{link1p6}.
\end{align}
Furthermore, invoking \cite[Equation (18)]{narayan2012}, we conclude that
\begin{align}
\lim_{\Delta\to 0}H(B_0^3|B_5)
&=\lim_{\Delta\to 0}H(B_0^3|B_5=1)=h(A_0^3)\label{link1p7}\\
\lim_{\Delta\to 0}H(B_0,B_2,B_3|B_6)&=\lim_{\Delta\to 0}H(B_0,B_2,B_3|B_6=1)\\
&=h(A_0,A_2,A_3)\label{link1p8}.
\end{align}
The proof of \eqref{link1toprove} is complete by invoking \eqref{link1part4} and \eqref{link1p5} to \eqref{link1p8}.

\bibliographystyle{IEEEtran}
\bibliography{IEEEfull_lin}

\end{document}

%% file: preamble.tex
\usepackage[mathscr]{eucal}
\usepackage[cmex10]{amsmath}
\usepackage{epsfig,epsf,psfrag}
\usepackage{amssymb,amsmath,amsthm,amsfonts,latexsym}
\usepackage{amsmath,graphicx,bm,xcolor,url,overpic}
\usepackage{fixltx2e}
\usepackage{array}
\usepackage{verbatim}
\usepackage{bm}
\usepackage{algorithmic}
\usepackage{algorithm}
\usepackage{verbatim}
\usepackage{textcomp}
\usepackage{mathrsfs}
\usepackage{epstopdf}

\newcommand{\openone}{\leavevmode\hbox{\small1\normalsize\kern-.33em1}}

\catcode`~=11 \def\UrlSpecials{\do\~{\kern -.15em\lower .7ex\hbox{~}\kern .04em}} \catcode`~=13 

\allowdisplaybreaks[4]

\newcommand{\nn}{\nonumber}

\newcommand{\calA}{\mathcal{A}}
\newcommand{\calB}{\mathcal{B}}
\newcommand{\calC}{\mathcal{C}}

\newcommand{\calE}{\mathcal{E}}

\newcommand{\calH}{\mathcal{H}}

\newcommand{\calK}{\mathcal{K}}

\newcommand{\calM}{\mathcal{M}}

\newcommand{\calR}{\mathcal{R}}
\newcommand{\calS}{\mathcal{S}}
\newcommand{\calT}{\mathcal{T}}
\newcommand{\calU}{\mathcal{U}}

\newcommand{\calW}{\mathcal{W}}
\newcommand{\calX}{\mathcal{X}}

\newcommand{\bF}{\mathbf{F}}

\newcommand{\rmA}{\mathrm{A}}

\newcommand{\rmB}{\mathrm{B}}
\newcommand{\rmc}{\mathrm{c}}
\newcommand{\rmC}{\mathrm{C}}
\newcommand{\rmd}{\mathrm{d}}

\newcommand{\rmH}{\mathrm{H}}

\newcommand{\rmP}{\mathrm{P}}

\newcommand{\rmU}{\mathrm{U}}

\newcommand{\bbN}{\mathbb{N}}

\newcommand{\bbR}{\mathbb{R}}

\DeclareMathAlphabet{\mathbsf}{OT1}{cmss}{bx}{n}
\DeclareMathAlphabet{\mathssf}{OT1}{cmss}{m}{sl}

\DeclareSymbolFont{bsfletters}{OT1}{cmss}{bx}{n}  
\DeclareSymbolFont{ssfletters}{OT1}{cmss}{m}{n}
\DeclareMathSymbol{\bsfGamma}{0}{bsfletters}{'000}
\DeclareMathSymbol{\ssfGamma}{0}{ssfletters}{'000}
\DeclareMathSymbol{\bsfDelta}{0}{bsfletters}{'001}
\DeclareMathSymbol{\ssfDelta}{0}{ssfletters}{'001}
\DeclareMathSymbol{\bsfTheta}{0}{bsfletters}{'002}
\DeclareMathSymbol{\ssfTheta}{0}{ssfletters}{'002}
\DeclareMathSymbol{\bsfLambda}{0}{bsfletters}{'003}
\DeclareMathSymbol{\ssfLambda}{0}{ssfletters}{'003}
\DeclareMathSymbol{\bsfXi}{0}{bsfletters}{'004}
\DeclareMathSymbol{\ssfXi}{0}{ssfletters}{'004}
\DeclareMathSymbol{\bsfPi}{0}{bsfletters}{'005}
\DeclareMathSymbol{\ssfPi}{0}{ssfletters}{'005}
\DeclareMathSymbol{\bsfSigma}{0}{bsfletters}{'006}
\DeclareMathSymbol{\ssfSigma}{0}{ssfletters}{'006}
\DeclareMathSymbol{\bsfUpsilon}{0}{bsfletters}{'007}
\DeclareMathSymbol{\ssfUpsilon}{0}{ssfletters}{'007}
\DeclareMathSymbol{\bsfPhi}{0}{bsfletters}{'010}
\DeclareMathSymbol{\ssfPhi}{0}{ssfletters}{'010}
\DeclareMathSymbol{\bsfPsi}{0}{bsfletters}{'011}
\DeclareMathSymbol{\ssfPsi}{0}{ssfletters}{'011}
\DeclareMathSymbol{\bsfOmega}{0}{bsfletters}{'012}
\DeclareMathSymbol{\ssfOmega}{0}{ssfletters}{'012}

\newcommand{\hatB}{\hat{B}}
\newcommand{\tilb}{\tilde{b}}

\newcommand{\tilR}{\tilde{R}}

\newcommand{\bara}{\bar{a}}

\DeclareMathOperator*{\argmax}{arg\,max}

\newtheorem{theorem}{Theorem} 
\newtheorem{lemma}[theorem]{Lemma}

\newtheorem{corollary}[theorem]{Corollary}
\newtheorem{definition}{Definition}